\documentclass[superscriptaddress,reprint,amsmath,amssymb,aps,pra]{revtex4-2}
\usepackage{color}
\usepackage{adjustbox}
\usepackage{soul}
\usepackage{placeins}
\usepackage{graphicx}
\usepackage{epstopdf}
\usepackage{natbib}
\usepackage[percent]{overpic}
\usepackage{graphicx}
\usepackage{hyperref}
\usepackage{dcolumn}
\usepackage{enumitem}
\usepackage{bm}
\usepackage{dcolumn}
\newcolumntype{d}[1]{D{.}{.}{#1}}

\newcommand*{\centt}[1]{\multicolumn{1}{c}{#1}}

\newcommand{\DR}{\Delta\! R}
\newcommand{\br}{\bm{r}}
\newcommand{\bR}{\bm{R}}

\newcommand{\Fkt}[1]{\,\mathsf {#1}}
\def\openone{\leavevmode\hbox{\small1\kern-3.3pt\normalsize1}}

\ifx\Tr\renewcommand{\Tr}{\Fkt{Tr}} 
\else\newcommand{\Tr}{\Fkt{Tr}}
\fi

\usepackage{booktabs}
\usepackage{graphicx}
\usepackage{amsmath}
\usepackage{indentfirst}
\begin{document}
\title{Second virial coefficients for helium-4 and helium-3 from
     accurate relativistic interaction potential}
\author{P. Czachorowski}
\email[Corresponding author: ]{pczachor@chem.uw.edu.pl}
\affiliation{Faculty of Chemistry, University of Warsaw, Pasteura 1, 02-093 Warsaw, Poland}
\author{M. Przybytek}
\affiliation{Faculty of Chemistry, University of Warsaw, Pasteura 1, 02-093 Warsaw, Poland}
\author{M. Lesiuk}
\affiliation{Faculty of Chemistry, University of Warsaw, Pasteura 1, 02-093 Warsaw, Poland}
\author{M. Puchalski}
\affiliation{Faculty  of  Chemistry,  Adam  Mickiewicz  University,  Umultowska  89b,  61-614  Pozna\'{n},  Poland}
\author{B. Jeziorski}
\affiliation{Faculty of Chemistry, University of Warsaw, Pasteura 1, 02-093 Warsaw, Poland}
\date{\today}
\begin{abstract}
The second virial coefficient and the second acoustic virial coefficient for helium-4 and helium-3 
are computed for a wide range of  temperatures ($0.5$--$1000$K) using a highly accurate 
nonrelativistic interaction potential [M. Przybytek {\emph{et al.}}, Phys. Rev. Lett. {\bf 119}, 123401 
(2017)] and recalculated relativistic and quantum-electrodynamic components.
The effects of the long-range retardation and of the nonadiabatic coupling of the nuclear and 
electronic motion are also taken into account. The results of our calculations represent at least  
fivefold improvement in accuracy compared to the previous {\emph{ab initio}} work. 
The computed virial coefficients agree well with the most accurate recent measurements but have 
significantly smaller uncertainty.     
\end{abstract}
\maketitle

\section{Introduction}
The existence of physical systems which can be both measured and theoretically described with a good accuracy is invaluable to science.
Not only can they help to test the consistency of our understanding of nature, but they may also
allow for development of new experimental techniques.
For fundamental metrology, helium can be considered such a system.
In contrast to hydrogen, seemingly simpler to describe, helium atoms interact particularly weakly.
For example, the helium dimer is either weakly bound in its single vibrational state (helium-4) or bound
states do not exist at all (helium-3).
This allows for accurate calculation of properties of gaseous helium solely in terms of pair 
interaction potential, as the three- and more-body effects become important only at larger 
pressures.
Due to the simplicity  of the system, theoretical description of the helium pair potential can  
include contributions beyond the Born-Oppenheimer nonrelativistic approximation.
Adiabatic, nonadiabatic, relativistic and quantum-electrodynamic (QED) effects can be added in a 
systematic manner, if the need arises.
The potential can be then used to predict properties of helium such as the energy of the 
 bound state or --- crucial in metrology --- the second density virial 
coefficient $B(T)$ and the acoustic virial coefficient  $\beta_a(T)$.
Accurate knowledge of these 
coefficients has been exploited by dielectric-constant gas thermometry \cite{Gaiser2015, 
Gaiser2019}, acoustic gas thermometry \cite{Moldover2014,Gavioso2019}, 
 single-pressure refractive-index gas thermometry (SPRIGT) \cite{Gao1, Gao}, and refractive-index gas thermometry \cite{Rourke2019}, as well as utilized in development of new pressure standards \cite{Schmidt07,Hendricks2018,Gaiser2020}.
At present, 
 the uncertainty of $B(T)$ dominates the uncertainty  budget for the electrical measurements 
of gas pressure at 7 MPa \cite{Gaiser2020} and for the SPRIGT measurements of temperature  below 20 K \cite{Gao}.
Thus, one can expect  that reducing the  error of the pair potential for helium and of the resulting virial coefficients will be of importance to experimental work in the field of thermal metrology.

The significance of the $B(T)$ coefficient  is best seen from the form of the virial equation of 
state, 
\begin{align}
p &=k_\mathrm{B}T\left[\rho+B(T)\rho^2 +  \cdots\right], \label{virial}
\end{align}
  used in accurate determination of the thermodynamic temperature  $T$, pressure 
$p$,  or, until 2019, the Boltzmann constant 
(fixed exactly at $k_\mathrm{B}$$=$$1.380649$$\times$$10^{-23}$~J/K
by the 2018 revision of the International System of
Units \cite{codata18}).  
The density $\rho$ can be determined from electrical measurements using the Clausius-Mossotti 
equation~\cite{ClausiusMossotti}
%
\begin{align}\label{CMeq}
\frac{\varepsilon_r-1}{\varepsilon_r+2}&=\frac{4\pi}{3}\alpha_\mathrm{d}\left[\rho+b_\varepsilon(T)\rho^2+  \cdots\right],
\end{align}
where    $\varepsilon_r$ is  the relative electric permittivity,   $\alpha_\mathrm{d}$ is the atomic   dipole polarizability,
and  $b_\varepsilon(T)$ is the second dielectric virial coefficient.
 Eliminating $\rho$ from Eqs.  (\ref{virial}) and (\ref{CMeq}), one can express the pressure through $\varepsilon_r$: 
\begin{align}\label{peq}
p&=\frac{k_\mathrm{B}T}{4\pi \alpha_\mathrm{d}  }(\varepsilon_r-1)\nonumber\\
&+\frac{k_\mathrm{B}T}{16\pi^2\alpha_\mathrm{d}^2}\!\left[B(T)\!- b_\varepsilon(T)\!-\!\frac{4\pi}{3}\alpha_\mathrm{d}\right]\!
(\varepsilon_r-1)^2+  \cdots,
\end{align}
with the  relative error of the order of $(\varepsilon_r-1)^2$. This error can be further reduced to $(\varepsilon_r-1)^3$ if the third virial coefficient $C(T)$ and  the third dielectric virial coefficient $c_\varepsilon(T)$  are included in Eqs. (\ref{virial}) and (\ref{CMeq}), respectively. 
An equation similar to Eq. (\ref{peq}) holds if $\rho$ is measured optically and $p$ is expressed via the index of refraction
  $n=(\varepsilon_r\mu_r)^{1/2}$,
$\mu_r$ being the magnetic permeability.  
The major difference would be that  the denominator in the first term of Eq. (\ref{peq}) would be  replaced by $2\pi (\alpha_\mathrm{d} +\chi_\mathrm{m})$,
where $\chi_\mathrm{m}$ is the magnetic susceptibility related to $\mu_r$ via $\mu_r=1 +4\pi \chi_\mathrm{m} \rho$.
Since $\chi_\mathrm{m}$ is much smaller than $\alpha_\mathrm{d}$ and $b_\varepsilon(T)$ is significantly smaller than $B(T)$, the major factors determining the accuracy of Eq. (\ref{peq}) and its optical variant are the accuracy of the polarizability 
$ \alpha_\mathrm{d}$  and  of the second virial coefficient $B(T)$.

In this paper we report accurate theoretical determination  of  $B(T)$ and  of the second 
acoustic virial coefficient $\beta_\mathrm{a}(T)$  for gaseous helium-4 and helium-3 within the wide 
temperature range $0.5$ -- $1000$ K.
We also present an improved helium pair interaction potential which can be used in  
calculations of other  thermophysical properties of gaseous helium or properties of the helium dimer 
itself. 

\section{Theory}
\subsection{Second virial coefficient}
The second virial coefficient $B(T)$ can be conveniently  expressed in terms of one- and two-atomic partition functions $Z_1$ and $Z_2$ \cite{Kilpaorig}
\begin{align}\label{B1}
B(T)&=- \mathcal{V}(Z_2-\frac{1}{2}Z_1^2)/Z_1^2,
\end{align}
where $\mathcal{V}$ is the volume of the system.
After inserting explicit forms of $Z_1$ and $Z_2$ \cite{Kilpatrick} into Eq.  (\ref{B1}), $B(T)$  can be partitioned into three   distinct  parts \cite{Hurly}:
\begin{align}
B(T)&=B_\mathrm{ideal}(T)+B_\mathrm{bound}(T)+B_\mathrm{th}(T), \label{secvir}
\end{align}
where $B_\mathrm{ideal}(T)$ is the  ideal gas   contribution, $B_\mathrm{bound}(T)$ is the effect contributed by bound rovibrational states of the dimer, and $B_\mathrm{th}(T)$ is the ``thermal  contribution'', dependent on the scattering dimer states.
For a monatomic, bosonic gas,  these contributions are defined as   \cite{Hurly}
\begin{align}
&B_\mathrm{ideal}(T)=-\frac{1}{16}\frac{1}{2s+1} \Lambda^3(T),\\
&B_\mathrm{bound}(T)=- \Lambda^3(T) \nonumber\\
&\times\Bigg\{\sum_{v,\,l}^{l\,\mathrm{even}}(2l+1)\frac{s+1}{2s+1}\left(e^{-E_{v,l}/(k_\mathrm{B}T)} - 1\right)\nonumber\\
&+\sum_{v,\,l}^{l\,\mathrm{odd}}(2l+1)\frac{s}{2s+1}\left(e^{-E_{v,l}/(k_\mathrm{B}T)} - 1\right)\Bigg\},\label{secvirbound}\\
&B_\mathrm{th}(T)=-\frac{ \Lambda^3(T)}{\pi k_\mathrm{B}T}\int_0^\infty dE \enskip e^{-E/(k_\mathrm{B}T)}\mathcal{S}(E),\label{secvirth}
\end{align}
where $v$ and $l$ in the summations in Eq. (\ref{secvirbound}) run over quantum numbers (vibrational 
and rotational, respectively) of the bound states of the system, $E_{v,l}$ are energies of these 
bound states, and $s$ is the spin of the nucleus.
Furthermore,
\begin{align}
\Lambda(T)&=\sqrt{2}\lambda_\mathrm{B}=\frac{h}{\sqrt{2\pi \mu_\mathrm{a}k_\mathrm{B}T}},\label{lambda}\\
\mathcal{S}(E)&=\sum_{l\,\mathrm{even}}^\infty(2l+1)\frac{s+1}{2s+1}\delta_l(E)\nonumber\\
&+\sum_{l\,\mathrm{odd}}^\infty(2l+1)\frac{s}{2s+1}\delta_l(E), \label{ssum}
\end{align}
where $\lambda_\mathrm{B}$ is the thermal de Broglie wavelength, $h$ is the Planck constant, 
$\mu_\mathrm{a}=m_\mathrm{a}/2$ is the reduced mass of two atoms, and $\delta_l(E)$ are phase shifts for the energy $E$ and the 
angular momentum quantum number $l$.
For helium-4, $s=0$ and there is one bound state, $E_{0,0}$, which has to be taken into account in Eq. (\ref{secvirbound}).
For fermionic gases, such as $^3$He ($s=1/2$), $B_\mathrm{ideal}$ changes sign, and the 
$s$-dependent prefactors (spin weights)  in the expressions for  $B_\mathrm{bound}(T)$ and  for 
$\mathcal{S}(E)$ are interchanged \cite{Kilpatrick}.
There are no bound rovibrational states of the helium-3 dimer, so for that isotope one has 
$B_\mathrm{bound}(T)=0$.

With the values of $B(T)$ calculated for a suitable range of temperatures $T$, one can easily 
obtain the second acoustic virial coefficient $\beta_\mathrm{a}(T)$ \cite{Hurly}:
\begin{align}
\beta_\mathrm{a}(T)&=2B(T)+2(\gamma_0-1)T\frac{dB(T)}{dT}\nonumber\\
&+\frac{(\gamma_0-1)^2}{\gamma_0}T^2\frac{d^2B(T)}{dT^2},\label{acvir}
\end{align}
where $\gamma_0$ is the heat capacity ratio ($\gamma_0=5/3$ for a monatomic gas with three degrees 
of freedom).
The differentiation of $B(T)$ is straightforward and can be done analytically using Eqs. 
(\ref{secvir})--(\ref{ssum}).

\subsection{Schr\"odinger equation}
To calculate $B_\mathrm{bound}(T)$ and $B_\mathrm{th}(T)$, the Schr\"odinger equation for the dimer  has to be solved   to determine the energy of the bound state of $^4$He$_2$ and the phase shifts.
In the center-of-mass frame, with the origin in the geometric center of the nuclei,  the 
equation for a binuclear molecule takes the form
\begin{align}
(H_\mathrm{el}+H_\mathrm{n}-E)\Psi(\br,\bR)&=0,
\end{align}
where
\begin{align}
H_\mathrm{el}&=-\frac{1}{2}\sum_i {\nabla}_{\br_i}^2-\sum_i\frac{Z_\mathrm{A}}{r_{i\mathrm{A}}}-\sum_i\frac{Z_\mathrm{B}}{r_{i\mathrm{B}}}+\sum_{i<j}\frac{1}{r_{ij}}+\frac{Z_\mathrm{A}Z_\mathrm{B}}{R},\\
H_\mathrm{n}&=-\frac{1}{2\mu_\mathrm{n}}\left( {\nabla}_{\bR}^2+ {\nabla}_\mathrm{el}^2\right)+\left(\frac{1}{M_\mathrm{A}}-\frac{1}{M_\mathrm{B}}\right) {\nabla}_\mathrm{el} {\nabla}_{\bR},\label{haen}
\end{align}
where the indices $i$, $j$ denote the electrons,  $ {\nabla}_\mathrm{el}=\frac{1}{2}\sum_i {\nabla}_{\br_i}$, \ $\br$ denote all electronic 
coordinates $\br_i$ collectively, A and B denote the nuclei, $M_\mathrm{A}$ and $M_\mathrm{B}$ are 
the nuclear masses, $\mu_\mathrm{n}=M_\mathrm{A}M_\mathrm{B}/(M_\mathrm{A}+M_\mathrm{B})$ is the 
reduced nuclear mass, $Z_\mathrm{A}$ and $Z_\mathrm{B}$ are the nuclear charges, 
and $\bR=\bR_\mathrm{A}-\bR_\mathrm{B}$ is a vector connecting the nuclei.
We are interested in homonuclear molecules only, so the rightmost term in Eq. (\ref{haen}) vanishes.
Note that in this section, as well as further on, we use atomic units (reduced Planck constant 
$\hbar$, elementary charge $e$, Bohr radius $a_0$, and electron mass $m_\mathrm{e}$ are used as units of action, electric charge, length, and mass, respectively), 
unless stated otherwise.

To describe the  finite-nuclear-mass effects, we employed the nonadiabatic perturbation theory (NAPT)  of Pachucki and Komasa \cite{Pachucki:08,PK2015}.
It assumes the wave function in the form
\begin{align}
 \Psi(\br,\bR)&\! =\!\psi(\br;\bR)\,Y_l^m(\theta,\phi)\chi_{l}(R)/R+\delta\Psi_\mathrm{na}(\br,\bR),
\end{align}
where $Y_l^m(\theta,\phi)$ is a spherical harmonic. 
It is assumed that $\langle\delta\Psi_\mathrm{na}|\psi\rangle_\mathrm{el}$=$0$, where the symbol $\langle 
\cdot|\cdot\rangle_\mathrm{el}$ denotes integration over the electronic coordinates only.
The function $\psi(\br;\bR)$ depends parametrically on $\bR$ and is a 
solution of the electronic Schr\"odinger equation (with fixed nuclear positions)
\begin{align}
H_\mathrm{el}(\bR)\psi(\br;\bR)&=\mathcal{E}(R)\psi(\br;\bR).
\end{align}
In the leading NAPT order, the equation for the nuclear wave function $\chi_{l}(R)$ is
\begin{align}
&\left[-\frac{1}{2\mu_\mathrm{n}}\frac{d^2}{d R^2} +\frac{l(l+1)}{2\mu_\mathrm{n} R^2} + V(R) -E\right]\chi_{l}(R)=0. \label{BOeq}
\end{align}
The potential $V(R)$ depends on the level of theory.
In the simplest case, in  the  nonrelativistic Born-Oppenheimer (BO) approximation, $V(R)=V_\mathrm{BO}(R)\equiv\mathcal{E}(R)$.
A detailed description of the potential used in our calculation is given in the next section.

While the adiabatic, relativistic, and QED corrections can be taken into account just by including a proper term in $V(R)$, the nonadiabatic effects require further modifications of Eq. (\ref{BOeq}) itself. When the finite-nuclear-mass effects up to $(m_\mathrm{e}/\mu_\mathrm{n})^2$ order are included in the Hamiltonian, the nuclear wave function $\chi_l(R)$ can be evaluated from \cite{PK2015}
\begin{align}
&\Bigg[-\frac{1}{2\mu_\parallel(R)}\frac{d^2}{d R^2}-\frac{d\mathcal{W}_{\parallel}(R)}{dR}\frac{d}{d R}+\frac{d\mathcal{W}_{\parallel}(R)}{dR}\frac{1}{R}\nonumber \\
&\,\,\,+\frac{l(l+1)}{2\mu_\perp(R)R^2}+V(R)+V_\mathrm{na}(R)-E\Bigg]\chi_l(R)=0, \label{NAPTeq}
\end{align}
with
\begin{align}
\frac{1}{2\mu_{\perp/\parallel}(R)}&=\frac{1}{2\mu_\mathrm{n}}+\mathcal{W}_{\perp/\parallel}(R),
\end{align}
where the potential $V(R)$ contains at least the BO and adiabatic contributions.
The most significant  change from Eq. (\ref{BOeq}) is the appearance of the ``distance-dependent masses'' in place of the reduced nuclear mass.
The functions $\mathcal{W}_{\parallel}(R)$, $\mathcal{W}_{\perp}(R)$, and $V_\mathrm{na}(R)$ take nonadiabatic effects into account and are defined in Ref. \cite{PK2015},  the last one denoted there as $\delta\mathcal{E}_\mathrm{na}(R)$. 
In fact, it is more convenient to use them with their value for the separated atoms limit subtracted \cite{PK2015,Przybytek:17}. Then
\begin{align}
\frac{1}{2\mu_{\perp/\parallel}(R)}&=\frac{1}{2\mu_\mathrm{a}}+\mathcal{W}^\mathrm{int}_{\perp/\parallel}(R)+O\left(\frac{1}{\mu^3_\mathrm{n}}\right),\\
\mathcal{W}^\mathrm{int}_{\perp/\parallel}(R)&=\mathcal{W}_{\perp/\parallel}(R)-\mathcal{W}_{\perp/\parallel}(\infty),\\
V^\mathrm{int}_\mathrm{na}(R)&=V_\mathrm{na}(R)-V_\mathrm{na}(\infty),
\end{align}
where $\mu_\mathrm{a}=\mu_\mathrm{n}+1$.
For the specific case of He$_2$, these functions were calculated in Ref.~\cite{Przybytek:17}. 

\subsection{Phase shifts}
As the internuclear distance $R$ increases, the wave function of the interacting atoms approaches 
that of free particles and can be written as \cite{Joachain}
\begin{align}
\chi_l(R)&\sim R\big[j_l(kR)-n_l(kR)\,\tan  \delta_{l}(E)\big],
\end{align}
where $k=\sqrt{2\mu_\mathrm{a}E}$ and the symbols $j_l(x)$ and $n_l(x)$ denote the spherical Bessel and Neumann functions, 
respectively.
By employing this expression, the phase shifts can be calculated as \cite{Joachain}
\begin{align}
\delta_l(E)&=\lim_{R\to\infty}\delta_{l}(E,R),\label{shifteq1}\\
\tan \delta_{l}(E,R)&=\frac{kj'_l(kR)-\gamma_{l}(R)j_l(kR)}{kn'_l(kR)-\gamma_{l}(R)n_l(kR)}\label{shifteq},\\
\gamma_{l}(R)&=\frac{\chi'_l(R)}{\chi_l(R)}-\frac{1}{R}. \label{shifteq3}
\end{align}
From a numerical standpoint, the above procedure can be repeated for increasing $R$ and stopped 
when the apparent shift function $\delta_{l}(E,R)$ converges to an acceptable level.


Note that Eq. (\ref{shifteq}) allows to calculate the phase shift only up to a multiple of $\pi$.
There are different methods to give it an absolute value, for example by defining the shift for zero or infinite energy 
and exploiting its continuity to obtain it for other energies, leaning on semiclassical expressions, or even
employing an alternative formula which provides an intrinsically absolute shift (compare, e.g., Ref. \cite{Chrysos}).
In our calculations, we require that $\delta_{l}(E)=0$ for a free particle [$V(R)\equiv0$].
As a result, for a repulsive potential one has $\delta_{l}(E)<0$, whereas for an attractive 
interaction $\delta_{l}(E)>0$ \cite{Joachain}.
Moreover, when $E\to 0$ for a fixed potential $V(R)$, the value of the phase shift tends to 
$n_l\pi$, where $n_l$ is the number of bound states supported by the potential for given angular 
momentum quantum number $l$ --- the behavior known from the Levinson theorem 
\cite{Levinson49,Joachain}.
Additionally, we require that the apparent shift $\delta_{l}(E,0)=0$ and then ensure that it is a continuous
function of $R$ during calculation \cite{Chrysos}.


\section{Pair potential for the helium dimer} \label{pots}
Following the approach used previously in
Refs.~\cite{Cencek:12} and \cite{Przybytek:17},
we represent the interaction energy of a pair of helium atoms 
as a sum of the BO potential, $V_\mathrm{BO}(R)$,
and a set of corrections that account for major post-BO effects:
the leading-order coupling of the nuclear and electronic
motion,   known as the adiabatic correction, 
$V_\mathrm{ad}(R)$,
the relativistic, $V_\mathrm{rel}(R)$, and QED effects, $V_\mathrm{QED}(R)$:
\begin{equation}
  \label{defV}
  V(R) = V_\mathrm{BO}(R)+V_\mathrm{ad}(R)+V_\mathrm{rel}(R)+V_\mathrm{QED}(R).
\end{equation}
All components of $V(R)$ for a given internuclear distance $R$
were obtained using the supermolecular approach by computing the difference between the 
respective dimer and atomic contributions
\begin{equation}
  \label{defVY}
  V_\mathrm{Y}(R)=\Delta E_\mathrm{Y} = E_\mathrm{Y}-2E_\mathrm{Y}^{A},
\end{equation}
where Y $=$ BO, ad, rel, QED.
$E_\mathrm{BO}$ is the nonrelativistic BO energy of the helium dimer.
$E_\mathrm{ad}$ is formally defined as the expectation value of
the nuclear kinetic energy operator \cite{Kutzelnigg:97}.
Therefore,
calculation of $E_\mathrm{ad}$ requires differentiation of
the clamped-nuclei wave function of the dimer
with respect to nuclear coordinates, on which the wave function depends
parametrically. $E_\mathrm{ad}$ can be calculated 
using various methods \cite{Handy:86,Ioannou:96,Handy:96,gauss06,gauss07,Pachucki:08}.
In Refs.~\cite{Przybytek:10,Cencek:12}, the Born-Handy approach \cite{Handy:86,Ioannou:96,Handy:96} was used,
while the results of Ref.~\cite{Przybytek:17} were obtained with the method proposed by Pachucki and Komasa \cite{Pachucki:08}.
$E_\mathrm{Y}$, Y $=$ rel, QED, are formally defined as expectation values of the operators
$\hat{H}_\mathrm{Y}$, shown below,
corresponding to a particular physical effect and computed
with the nonrelativistic electronic BO function.
The atomic contributions, $E_\mathrm{Y}^{A}$, Y $=$ BO, ad, rel, QED,
are defined similarly, but correspond to a single helium atom.

The operator $\hat{H}_\mathrm{rel}$ is the Breit-Pauli Hamiltonian
\cite{BeSal},
which for closed-shell systems in a singlet state 
consists of the mass-velocity operator $\hat{H}_\mathrm{mv}$,
the one- and two-electron Darwin operators $\hat{H}_\mathrm{D1}$ and $\hat{H}_\mathrm{D2}$,
and the Breit operator $\hat{H}_\mathrm{Br}$:
\begin{equation}
  \label{Hrel}
  \hat{H}_\mathrm{rel} =
  \hat{H}_\mathrm{mv} +
  \hat{H}_\mathrm{D1} +
  \hat{H}_\mathrm{D2} +
  \hat{H}_\mathrm{Br},
\end{equation}
where
\begin{align}
  \label{Hmv}
  \hat{H}_\mathrm{mv}&=-\frac18\alpha^2\sum_i p_i^4, \\
  \label{HD1}
  \hat{H}_\mathrm{D1}&=\frac\pi2\alpha^2\sum_I\sum_iZ_I\delta(\bm{r}_i-\bm{r}_I),\\
  \label{HD2}
  \hat{H}_\mathrm{D2}&=\pi\alpha^2\sum_{i<j}\delta(\bm{r}_{ij}),  \\
  \label{HBr}
  \hat{H}_\mathrm{Br}&=-\frac12\alpha^2\sum_{i<j}
  \left[
    \frac{\bm{p}_i\cdot\bm{p}_j}{r_{ij}}+\frac{\bm{r}_{ij}\cdot(\bm{r}_{ij}\cdot\bm{p}_j)\bm{p}_i}{r_{ij}^3}
    \right].
\end{align}
In these equations, 
$\bm{r}_{ij}=\bm{r}_{i}-\bm{r}_{j}$, $\bm{p}_i=-\mathrm{i}\nabla_{\bm{r}_i}$,
the index $I$ runs over all the nuclei, with charge $Z_I$ and
located at the position $\bm{r}_I$,
  $\delta(\bm{r})$ is the Dirac delta function, and 
  $\alpha  = 1/137.035\,999\,084$   is the fine structure constant \cite{codata18}. 
%
The sum of one-electron operators is usually referred to as the Cowan-Griffin
(CG) operator \cite{Cowan:76},
$\hat{H}_\mathrm{CG}=\hat{H}_\mathrm{mv}+\hat{H}_\mathrm{D1}$.

The operator $\hat{H}_\mathrm{QED}$,  defining the QED correction, can be expressed as the    linear combination
\begin{equation}
  \label{HQED}
  \begin{split}
    \hat{H}_\mathrm{QED}
    &=\frac{8\alpha}{3\pi}\left(\frac{19}{30}-2\ln\alpha-\ln k_0\right) \hat{H}_\mathrm{D1}\\
    &+\frac\alpha\pi\left(\frac{164}{15}+\frac{14}3\ln\alpha\right)\hat{H}_\mathrm{D2}+\hat{H}_\mathrm{AS},
  \end{split}
\end{equation}
 of    $\hat{H}_\mathrm{D1}$,    $\hat{H}_\mathrm{D2}$ 
and the Araki-Sucher (AS)  operator $ \hat{H}_\mathrm{AS}$ 
\cite{Araki:57,Sucher:58,Pachucki:98} defined as 
 \begin{align}
    \label{HAS}
  \hat{H}_\mathrm{AS}&=-\frac7{6\pi}\alpha^3\sum_{i<j}\hat{P}(r_{ij}^{-3}),
\end{align}
where   $\hat{P}(r_{ij}^{-3})$ is  the operator distribution  
\begin{equation}
  \langle\hat{P}(r_{ij}^{-3})\rangle=\lim_{a\to 0}
  \langle r_{ij}^{-3}\theta(r_{ij}-a)+4\pi(\gamma+\ln a)\delta(\bm{r}_{ij})\rangle,
\end{equation}
with  $\theta(r) $ and $\gamma$ standing here for the Heaviside step function and
the Euler-Mascheroni constant, respectively.
The quantity  $\ln k_0$ in Eq. (\ref{HQED}) is the so-called Bethe logarithm \cite{BeSal}. 
 The interatomic distance dependence of $\ln k_0$ is weak \cite{Piszcz:09}, especially for a 
 weakly interacting system such as the helium dimer \cite{Cencek:12}, and can be neglected at the accuracy
 level considered in this work. Therefore, we  fixed  $\ln k_0$ at its helium atom value equal to
 $4.370\,160\,222\,0(1)$ \cite{Korobov:04}.

In the present work, the values of all four
components of $V(R)$ in Eq.~\eqref{defV} were obtained for a set of 55 distances 
ranging from 1 to 30 bohrs.
The recommended BO interaction energies for 46 distances, $1\leq R \leq 9$~bohrs,
and the adiabatic corrections for a full set of distances
were taken from Ref.~\cite{Przybytek:17}
together with their estimated theoretical uncertainties.
The remaining data points ---
the BO energies at nine distances, $10\leq R\leq 30$~bohrs,
and the relativistic and QED corrections --- were recalculated using a
Gaussian orbital approach with larger one-electron basis sets than the ones employed
in Ref.~\cite{Przybytek:17}.
The new calculations revealed that  
theoretical uncertainties of some components of  $V_\mathrm{rel}(R)$  estimated  in Ref.~\cite{Przybytek:17}
were too optimistic
so they were carefully reexamined in the present work.
The AS interaction energies were calculated initially in
Ref.~\cite{Cencek:12} (for 17 interatomic distances only) using explicitly correlated Gaussian (ECG)
expansion of the wave function for the dimer and
near exact atomic AS energy \cite{Drake:88}.
Due to the basis set superposition error (BSSE), they had absolute uncertainties of
virtually the same magnitude for all $R\geq5$~bohrs. 
As a result, the AS correction 
was the dominant source of theoretical error of the pair potential $V(R)$
for large distances. 

The BO energies (at $10\leq R\leq 30$~bohrs) and the expectation values
of the operators constituting the relativistic correction, 
Eqs.~\eqref{Hmv}--\eqref{HBr},
were evaluated at two levels of theory:
the coupled-cluster method with single, double, and noniterative triple
excitations [CCSD(T)], and the full configuration interaction (FCI) method.
Note that the calculations at the CCSD(T) level were performed utilizing the Hellman-Feynman theorem, according to the
linear response theory \cite{Coriani:04}.
The individual interaction energies were then obtained
using a two-step procedure as the following sum:
\begin{equation}
  \label{defCpF}
  \Delta E_\mathrm{Y}=\Delta E_\mathrm{Y}^\mathrm{CCSD(T)}+\delta E_\mathrm{Y}^\mathrm{FCI},
\end{equation}
where Y $=$ BO, mv, $D_1$, $D_2$, Br. 
The first, dominating  term   $\Delta E_\mathrm{Y}^\mathrm{CCSD(T)}$  is defined by
 Eq. (\ref{defVY}) and
the  much smaller FCI correction $\delta E_\mathrm{Y}^\mathrm{FCI}$  is defined as
\begin{equation}
  \label{defdF}
  \delta E_\mathrm{Y}^\mathrm{FCI}=\Delta E_\mathrm{Y}^\mathrm{FCI}-\Delta E_\mathrm{Y}^\mathrm{CCSD(T)}
\end{equation}
with the  quantities on the right-hand side of  Eq.~\eqref{defdF} computed  using the same basis set.

The AS correction, defined by the operator in Eq.~\eqref{HAS}, was
determined only at the FCI level:
\begin{equation}
  \Delta E_\mathrm{AS}=\Delta E_\mathrm{AS}^\mathrm{FCI}.
\end{equation}
When  computing $\Delta E_\mathrm{Y}^\mathrm{CCSD(T)}$ and
$\Delta E_\mathrm{Y}^\mathrm{FCI}$  via  Eq. 
  \eqref{defVY}, the atomic properties $E_\mathrm{Y}^A$
were obtained  with  the corresponding dimer-centered basis set,
which is equivalent to using the so-called counterpoise scheme,
which corrects for BSSE \cite{Boys:70}.
All calculations were performed using modified $dXZ$ basis sets of
Ref.~\cite{Cencek:12} (containing 21 uncontracted $s$ functions)
with the cardinal numbers $X$ up to $X=8$ for CCSD(T) and up to $X=7$ for FCI.
The largest FCI calculations employed a wave function with approximately
$2\times10^9$ determinants (at $D_\mathrm{2h}$ symmetry).
The CCSD(T) calculations were performed using
the \textsc{dalton} 2013 package \cite{daltonpaper,dalton13},
whereas at the FCI level we used a program  \cite{hector} written specifically for the 
purpose  of the present  project. In the latter case, the Hartree-Fock orbitals,
the standard one- and two-electron integrals, and integrals involving
the relativistic operators were generated using the local version of
the \textsc{dalton} 2.0 package \cite{daltonpaper,dalton2,Coriani:04},
while integrals involving the AS operator were computed using
the computer code developed in Ref.~\cite{Balcerzak:17}.

To reduce the basis set incompleteness errors in the quantities obtained from the Gaussian orbital calculations, we employed  the Riemann 
extrapolation scheme introduced in Ref.~\cite{lesiuk19}. This method assumes that the differences 
$\delta_X=E_X-E_{X-1}$, where $E_X$ are the quantities of interest calculated within a basis set of 
cardinal number $X$, behave asymptotically  as $\delta_X\sim \mbox{const}\times X^{-n}$ for 
$X\rightarrow\infty$. As shown in Ref.~\cite{lesiuk19}, this leads to the following two-point 
formula for the complete basis set (CBS) limit:
\begin{align}
 E_\infty = E_X + X^n\big(E_X-E_{X-1}\big)\Big[\zeta(n) - \sum_{i=1}^X i^{-n} \Big],
\end{align}
where $\zeta(s)=\sum_{i=1}^\infty i^{-s}$ is the Riemann zeta function. In the case of the 
Born-Oppenheimer potential,  we used $n=4$ in the above formula. The same scheme was found to be 
adequate for the mass-velocity and 
one-electron Darwin corrections. However, the remaining contributions to the total potential have a 
different convergence rate. In the case of the two-electron Darwin correction we employed
 $n=2$, in agreement with the analytic results of Kutzelnigg~\cite{kutz08} 
for the helium atom. The Breit correction was extrapolated with $n=5/2$ as suggested by the 
previous numerical results for He$_2$ \cite{Cencek:12} [using $\zeta(5/2)\approx1.34149$]. To 
apply the Riemann extrapolation to the AS correction, the method presented in 
Ref.~\cite{lesiuk19} has to be extended because the quantities $\delta_X$ behave asymptotically as 
$\delta_X\sim aX^{-2}\ln X+bX^{-2}$, where $a$ and $b$ are numerical constants (see 
Ref.~\cite{Balcerzak:17}). Combining results from three consecutive basis sets, we find the following 
expression for the CBS limit in this case:
\begin{align}
 E_\infty = E_X + a\Bigg[-\zeta'(2) - \sum_{i=1}^X i^{-2}\ln i\Bigg]
          + b\Bigg[ \zeta(2) - \sum_{i=1}^X i^{-2} \Bigg],
\end{align}
where $\zeta'(s)$ is the derivative of the Riemann zeta function $\big[\zeta'(2)\approx 
-0.937548\big]$. The constants $a$ and $b$ are found from the expressions
\begin{align}
 a &= \Bigg[ X^2 \big(E_X-E_{X-1}\big) - (X-1)^2 \big(E_{X-1}-E_{X-2}\big) \Bigg]\nonumber\\
&\times \Big[\ln(X) -\ln(X-1)\Big]^{-1}, \\
 b &= a\cdot\ln X - X^2 \big(E_X-E_{X-1}\big).
\end{align}
To demonstrate  the efficacy of the Riemann extrapolation of the AS correction, one  can 
make a comparison  with  the ECG results of Cencek {\emph{et al.}} \cite{Cencek:12}. 
At the distance $R=2.0\,$a.u. the ECG result $-0.032\,25(25)$ mK  is accurate enough to be treated 
as a reference. The values obtained using three largest Gaussian orbital basis sets ($X =5,6,7$) are 
equal to $-0.028\,50$, $-0.028\,88$, and $-0.029\,18\,$mK, respectively, and converge rather 
slowly. 
The Riemann extrapolation  yields $-0.032\,23\,$mK, in good agreement with the ECG 
value. In the attractive part of the potential, the ECG results became inaccurate due to the 
lack of a BSSE correction, so the Riemann-extrapolated results are much more 
accurate in this region. 

For each quantity considered in this work, the CCSD(T) results, 
$\Delta E_\mathrm{Y}^\mathrm{CCSD(T)}$, and the FCI correction, $\delta E_\mathrm{Y}^\mathrm{FCI}$, 
were extrapolated separately. The sole exception is the AS correction, where only the FCI 
results are available and thus were extrapolated directly. The errors of the extrapolated 
quantities were assigned as follows. In the case of the mass-velocity and one-electron Darwin 
corrections the extrapolation error is conservatively estimated as a difference between the 
extrapolated result and the value obtained in the largest basis set available.
For the remaining three corrections, this straightforward approach
is not adequate since it leads to a gross overestimation of error.
To 
circumvent this problem, we introduce a modified procedure where the difference between the 
extrapolated result and the value obtained in the largest basis set is scaled by a constant 
(independent of the internuclear distance) to get the error estimate. We found that the scaling by 
a factor of $0.1$ is adequate for the two-electron Darwin and AS corrections, 
while $0.5$ is used for the Breit correction. 
This approach is validated by comparing with the ECG results \cite{Cencek:12}
at short interatomic distances --- where the ECG approach is reliable and gives small uncertainties. 
 For example, at 
$R=2.0\,$a.u. we obtained $1.121(11)$, $1.944(5)$, and $-0.0322(3)$ for the two-electron Darwin, 
Breit, and AS corrections, respectively, after scaling the errors, while the 
corresponding ECG results read $1.132(5)$, $1.9411(1)$, and $-0.0322(3)$. 
Clearly, the two sets of results are in full agreement and a similar picture is obtained
for other interatomic distances where the ECG results are still reliable.
For all corrections other than the AS, the 
errors of the CCSD(T) and FCI contributions were added quadratically to obtain the final error 
estimates.

To ensure high accuracy of the pair potential
for large interatomic distances $R$, we recalculated the  constants
$C_n(\mathrm{Y})$, $n<8$, Y $=$ ad, rel, QED,
determining the leading-order terms
in the asymptotic expansion in powers of $1/R$
of the post-BO terms in Eq.~\eqref{defV},
$V_\mathrm{Y}(R)\sim-\sum_nC_n(\mathrm{Y})/R^n$.
The coefficients $C_n(\mathrm{rel})$ and $C_n(\mathrm{QED})$ are defined
as appropriate combinations of the coefficients
calculated separately for the components of
$\hat{H}_\mathrm{rel}$ and $\hat{H}_\mathrm{QED}$
according to Eqs.~\eqref{Hrel} and \eqref{HQED}.
Following Ref.~\cite{Cencek:12}, we distinguish between contributions
to $C_n(\mathrm{Y})$, Y $=$ mv, D1, D2, Br, AS, coming from
either the intra- or intermonomer part of a given operator $\hat{H}_\mathrm{Y}$.
The asymptotic expansion of
the intramonomer part always starts with the term proportional to $1/R^6$
and the corresponding coefficient expressed in the sum-over-states form is
\cite{Cencek:12}
\begin{equation}
  \label{Cintra} 
  \begin{split}
    C_6(Y,\mathrm{intra})=&-12\sum_{abc}\frac{Z_{0a}Z_{0b}Z_{0c}^2Y_{ab}}{(\omega_a +\omega_c)(\omega_b +\omega_c )} \\
    &-24\sum_{abc}\frac{Z_{0a}Z_{ab}Z_{0c}^2Y_{0b}}{(\omega_a +\omega_c)\omega_b }.
  \end{split}
\end{equation}
The matrix elements $Z_{ab}$ and $Y_{ab}$ are
\begin{align}
  Z_{ab}&=\langle\phi_a|\sum_{i=1}^2z_i|\phi_b\rangle, \\
  Y_{ab}&=\langle\phi_a|\hat{H}_\mathrm{Y}^{A}|\phi_b\rangle-
  \delta_{ab}\langle\phi_0|\hat{H}_\mathrm{Y}^{A}|\phi_0\rangle,
\end{align}
where $\hat{H}_\mathrm{Y}^A$ are the operators
from Eqs.~\eqref{Hmv}--\eqref{HBr} and \eqref{HAS} defined for a single helium atom,
$\phi_0$ is the ground-state wave function of helium, $\phi_a$ are
the wave functions of the excited states, and $\omega_a$ are the corresponding
excitation energies.
As it was shown in Ref.~\cite{Przybytek:12ad}, the leading-order coefficient
for the adiabatic correction, $C_6(\mathrm{ad})$, can be calculated using
the same formula, Eq.~\eqref{Cintra}, and in this case
the $\hat{H}_\mathrm{ad}^A$ operator has the form
\begin{equation}
  \hat{H}_\mathrm{ad}^A=\frac1{2M}\left(\sum_{i=1}^2\bm{p}_i\right)^2,
\end{equation}
where $M=7294.299\,541\,42(24)$ \cite{codata18}
is the mass of the helium-4 atom nucleus.
Only the two-electron operators, Eqs.~\eqref{HD2}, \eqref{HBr}, and \eqref{HAS}, have an 
intermonomer
part and, among them, only the Breit and AS operators give contributions
to the interaction energy that vanish for large distances
as powers of $1/R$ \cite{Cencek:12}. For the Breit interaction, the leading-order
coefficients are \cite{Cencek:12,Meath:66}
\begin{align}
  \label{C4Br}
  C_4(\mathrm{Br},\mathrm{inter})&=2\sum_{ab}\frac{Z_{0a}P_{0a}\,Z_{0b}P_{0b}}{\omega_a+\omega_b}, \\
  \label{C6Br}
  C_6(\mathrm{Br},\mathrm{inter})&=18\sum_{ab}\frac{Z_{0a}P_{0a}\,Q_{0b}S_{0b}}{\omega_a+\omega_b}\nonumber \\
&-\frac{12}5\sum_{ab}\frac{Z_{0a}P_{0a}\,Z_{0b}T_{0b}}{\omega_a+\omega_b},
\end{align}
where
\begin{align}
  Q_{0a}&=\langle\phi_0|\sum_{i=1}^2\frac12\left(3z_i^2-r_i^2\right)|\phi_a\rangle, \\
  P_{0a}&=\langle\phi_0|\sum_{i=1}^2p_{zi}|\phi_a\rangle, \\
  S_{0a}&=\langle\phi_0|\sum_{i=1}^2z_ip_{zi}|\phi_a\rangle, \\
  T_{0a}&=\langle\phi_0|\sum_{i=1}^2\left[2r_i^2p_{zi}-z_i(\bm{r}_i\cdot\bm{p}_i)\right]|\phi_a\rangle.
\end{align}
For the AS interaction, the odd-$n$ coefficients up to $n=7$ are
determined exclusively by the expectation value of the intermonomer part of
$\hat{H}_\mathrm{AS}$ calculated with the product of ground-state wave functions
of both interacting atoms. Utilizing multipole expansion of
the $r_{ij}^{-3}$ operator 
   \cite{Sack:64},
it is easy to show that
\begin{align}
  C_3(\mathrm{AS},\mathrm{inter})&=\frac7{6\pi}\alpha^3R_0^2, \\
  C_5(\mathrm{AS},\mathrm{inter})&=\frac7{3\pi}\alpha^3R_0\,R_1, \\
  C_7(\mathrm{AS},\mathrm{inter})&=\frac7{9\pi}\alpha^3\left(3R_0\,R_2+5R_1^2\right),
\end{align}
where
\begin{equation}
  \label{RnAS}
  R_n=\langle\phi_0|\sum_{i=1}^2r_i^{2n}|\phi_0\rangle.
\end{equation}
  The $C_n$ and $R_n$ coefficients,
Eqs.~\eqref{Cintra}, \eqref{C4Br}, \eqref{C6Br}, and \eqref{RnAS},
were calculated using ECG expansions of the wave functions of a helium atom
with $N_\mathrm{b}=128$, $256$, and $512$ terms for $\phi_0$ and $2N_\mathrm{b}$
terms for the excited states $\phi_a$ in intermediate summations.
The values presented in Table~\ref{table:Cn} were obtained
by taking the results calculated with $N_\mathrm{b}=512$ as
the recommended values with their uncertainties estimated as
the absolute difference between the $N_\mathrm{b}=256$ and $N_\mathrm{b}=512$
results. The reliability of this procedure was checked by computing
the leading-order coefficient in the asymptotic expansion of
the $V_\mathrm{BO}$ potential. The obtained value
$C_6(\mathrm{BO})=1.460\,977\,837\,723\,6(2)$ agrees to all significant digits
with the value $C_6(\mathrm{BO})=1.460\,977\,837\,725(2)$
taken from the literature \cite{Zhang:06} but is somewhat more accurate.
The final results for $C_n(\mathrm{Y})$, $n<8$, Y $=$ ad, rel, QED,
calculated using data from Table~\ref{table:Cn} agree with the ones used
in Ref.~\cite{Cencek:12} but have two to three times more significant digits.

\begin{table}
  \caption{
    \label{table:Cn}
    Components of the leading asymptotic constants of
    $V_\mathrm{ad}(R)$, $V_\mathrm{rel}(R)$, and $V_\mathrm{QED}(R)$.
    The labels ``intra'' and ``inter'' were omitted, when a given $C_n$ has
    only one contribution of either type.}
  \begin{ruledtabular}
    \begin{tabular}{cd{17}}
      $2M_N\, C_6(\mathrm{ad})$ & 16.699\,662\,17(5) \\
      $\alpha^{-2}\, C_6(\mathrm{mv})$ & -31.628\,828(6) \\
      $\alpha^{-2}\, C_6(\mathrm{D1})$ & 26.786\,047(3) \\
      $\alpha^{-2}\, C_6(\mathrm{D2})$ & 1.934\,254\,7(6) \\
      $\alpha^{-2}\, C_4(\mathrm{Br})$ & -0.663\,309\,369\,557\,98(6) \\
      $\alpha^{-2}\, C_6(\mathrm{Br,intra})$ & -0.954\,535\,671(6) \\
      $\alpha^{-2}\, C_6(\mathrm{Br,inter})$ & -2.603\,188\,510\,963(6) \\
      $\alpha^{-3}\, C_6(\mathrm{AS})$ & -1.409\,909(1) \\
      $R_1$ & 2.386\,965\,990\,037\,8(1) \\
      $R_2$ & 7.947\,129\,863\,325(1) \\
    \end{tabular}
  \end{ruledtabular}
\end{table}

The extrapolated values of
$V_\mathrm{BO}(R)$, $V_\mathrm{ad}(R)$, $V_\mathrm{rel}(R)$, and $V_\mathrm{QED}(R)$
were fitted separately to the analytic functions of the form
\begin{equation}
  \sum_{k=1}^M e^{-a_k R}\sum_{i=I_0}^{I_1}P_{ik}R^i
  -\sum_{n=N_0}^{N_1}f_n(\eta R)\frac{C_n}{R^n},
\end{equation}
where $f_n(x)=1-e^{-x}\left(\sum_{i=0}^{n}x^i/i!\right)$
is the Tang-Toennies damping function \cite{Tang:84},
$a_k$, $P_{ik}$, and $\eta$ are adjustable parameters,
and the summation limits $[M,I_0,I_1,N_0,N_1]$ are
$[3,-1,2,6,16]$ for $V_\mathrm{BO}(R)$,
$[3, 0,2,6,10]$ for $V_\mathrm{ad}(R)$,
$[2, 0,2,4,10]$ for $V_\mathrm{rel}(R)$, and
$[3, 0,2,3,10]$ for $V_\mathrm{QED}(R)$.
The asymptotic constants $C_8$ and $C_{10}$ for
$V_\mathrm{rel}(R)$ and $V_\mathrm{QED}(R)$ are not known and were   adjusted.
In both cases $C_9$ was neglected.
The remaining constants $C_n$ were fixed and set equal to the values known
from the literature \cite{Zhang:06,Tang:11,Przybytek:08,Przybytek:12ad}
or to the values calculated in this work as described above.
In the analytical fitting of $V_\mathrm{BO}(R)$,
the linear parameters $P_{ik}$ were constrained by imposing the condition
\begin{equation}
  V_\mathrm{BO}(R)=\frac4R+(E_\mathrm{Be}-2E_\mathrm{He})+O(R^2)
\end{equation}
that ensures the correct short-range asymptotics of the potential.
The known accurate ground-state energies of the beryllium and helium atoms,
$E_\mathrm{Be}=-14.667356498$ \cite{Puchalski:13} and
$E_\mathrm{He}=-2.903724377$ \cite{Handbook}, were used.
Similarly, the analytical fits of the post-BO corrections were constrained
to ensure correct values of the potentials at $R=0$.
The corresponding conditions,
$V_\mathrm{ad}(0)=0.0001971680204$,
$V_\mathrm{rel}(0)=-0.00215235927$, and
$V_\mathrm{QED}(0)=0.00039644284$,
were obtained using data from Ref.~\cite{Puchalski:13} for
the berylliumlike united atom, and from Refs.~\cite{Handbook,Drake:88}
for the helium atom.
In all cases, the inverse squares of the uncertainties $\sigma$ were
used as the weighting factors to ensure that the fit accuracy is higher
in regions of more accurate data points. The average absolute errors of
the fits are
0.18$\sigma$ for $V_\mathrm{BO}(R)$,
0.13$\sigma$ for $V_\mathrm{ad}(R)$,
0.15$\sigma$ for $V_\mathrm{rel}(R)$, and
0.13$\sigma$ for $V_\mathrm{QED}(R)$.
In some cases,
the fitted data points are reproduced with errors that are greater than
the estimated data point uncertainties. This behavior was observed only for
$R=26$ and $30$ bohrs. For such large distances
the values of the potentials are small and their an accurate prediction
using the supermolecular approach is difficult due to large cancellation
of significant digits between the dimer and atomic contributions
in Eq.~\eqref{defVY}. On the other hand, in this region the potentials
are entirely determined by their asymptotic expansion. Therefore,
the analytic functions that include accurate asymptotic constant $C_n$
are expected to provide more reliable results than the ones calculated
from Eq.~\eqref{defVY}.

In order to estimate the uncertainties of physical properties of helium
calculated with the present potential, we constructed functions
$\sigma_\mathrm{Y}(R)$, Y $=$ BO, ad, rel, QED,
representing estimated uncertainties
of the components $V_\mathrm{Y}(R)$ of the interaction potential,
such that the exact values of a given component can be assumed
to be contained between functions $V_\mathrm{Y}(R)\pm \sigma_\mathrm{Y}(R)$.
The functions $\sigma_\mathrm{Y}(R)$ are not intended to accurately reproduce
the estimated uncertainties but to follow general trends
in their $R$ dependence and to bound most values from above.
Analytic functions $\sigma_\mathrm{Y}(R)$
used to represent the uncertainties have the general form
\begin{equation}
  s_0e^{-a_0R}+\sum_{i=1}^{n}s_ie^{-a_iR^2},
\end{equation}
where $a_i$ and $s_i$  are adjustable parameters, and the summation limit $n$ is
4 for $\sigma_\mathrm{BO}(R)$,
and 3 for $\sigma_\mathrm{ad}(R)$,
$\sigma_\mathrm{rel}(R)$, and $\sigma_\mathrm{QED}(R)$.
The fit of uncertainties was performed using the standard least-squares method
applied to a reduced set of data points obtained by discarding points where
the values of uncertainties are significantly smaller than the neighboring ones.
The value of $a_0$ was adjusted only once, while constructing the function
$\sigma_\mathrm{BO}(R)$, and then set fixed during generation of the remaining
functions. The average ratio of the value of $\sigma_\mathrm{Y}(R)$
to the value of estimated uncertainty
calculated for a whole set of 55 distances is
1.33 for $\sigma_\mathrm{BO}(R)$,
1.81 for $\sigma_\mathrm{ad}(R)$,
1.00 for $\sigma_\mathrm{rel}(R)$, and
1.04 for $\sigma_\mathrm{QED}(R)$.

The values of all parameters of the functions $V_\mathrm{Y}(R)$ and
$\sigma_\mathrm{Y}(R)$ (Y $=$ BO, ad, rel, QED),
and a numerical implementation of the fits in the form of a \textsc{fortran} 2003 code
can be found in the Supplemental Material \cite{supp}.


The effects of retardation (see Ref. \cite{Przybytek:12cp} for their precise definition), were included in
the potential $V(R)$ using the procedure employed in Ref.~\cite{Cencek:12}. In the retardation
damping function $g(x)=(1+\sum_{n=1}^5 A_n x^n)/(1+\sum_{n=1}^6 B_n x^n)$, the coefficients
$B_n$ were taken from Ref.~\cite{Cencek:12}, while $A_n$ were recalculated using Eqs. ($48$) -- ($52$) from the same reference, to conform to more accurate
values of the asymptotic constants calculated in the present work.

To conclude this section, let us summarize
the improvements that have been made
in the description of the helium pair potential since 2012 \cite{Cencek:12}.
The most important achievement is a consistent reduction of errors of
the dominant BO component for all distances by about one order of magnitude
(4--23 times)
done in Ref.~\cite{Przybytek:17}. In the case of post-BO corrections,
the estimated uncertainties of the present potentials are similar
to the ones from Ref.~\cite{Cencek:12}
in the highly repulsive region of $V(R)$ for $R\leq3$ and smaller for larger
distances. This reduction is by one to two orders of magnitude (7--222 times)
for the adiabatic
correction and by a factor of about 5 for the relativistic correction.
The most significant changes are observed for the QED components where,
due to a proper removal of BSSE in the AS term, the ratio of errors
estimated in Ref.~\cite{Cencek:12} to the present ones grows steeply
from 1.4 at $R=3.5$ to $5\times10^3$ at $R=12$.
Besides reducing the theoretical errors, we were also able to calculate
the potential on a much finer grid of points
(55 compared to 17 in Ref.~\cite{Cencek:12}) and to improve the description
of the long-range decay of post-BO corrections where both
the number of terms included in their asymptotic expansion in powers of $1/R$
and the precision of the asymptotic constants were increased.
All these factors combined allowed us to produce a more robust and reliable
analytical representation of $V(R)$ and its uncertainties $\sigma(R)$ that
are needed in the determination of thermophysical properties of helium. 

\section{Numerical calculations of second virial coefficient}
The nonadiabatic nuclear Schr\"odinger equation (\ref{NAPTeq}) has the form
\begin{align}
&\left[\frac{d^2}{d R^2}+p(R)\frac{d}{d R}+q(R)\right]f(R)=0 \label{numeq2},
\end{align}
with
\begin{align}
f(R)&\equiv\chi_l(R),\\
p(R)&\equiv2\mu_\parallel(R)\frac{d\mathcal{W}^\mathrm{int}_\parallel(R)}{dR},\\
q(R)&\equiv2\mu_\parallel(R)\Bigg[E-V(R)-V^\mathrm{int}_\mathrm{na}(R)\nonumber \\
  &-\frac{1}{R}\frac{d\mathcal{W}^\mathrm{int}_\parallel(R)}{dR}
  -\frac{l(l+1)}{2\mu_\perp(R)R^2}\Bigg].
\end{align}
A standard approach to finding a solution of such an equation is the Numerov method \cite{Numerovorig,Blatt}.
However, in the standard formulation of this method there is no first derivative present in the 
equation.
To cast Eq. (\ref{numeq2}) into the required form, a substitution can be used \cite{Leroy} to 
remove the problematic first-derivative term
\begin{align}
f(R)&=\phi(R)e^{-\int dR\, p(R)/2}.
\end{align}
This leads to the equation
\begin{align}
&\left[\frac{d^2}{d R^2}+Q(R)\right]\phi(R)=0,
\end{align}
where
\begin{align}
Q(R)&\equiv q(R)-\frac{1}{4}p^2(R)-\frac{1}{2}\frac{dp(R)}{dR},
\end{align}
which can now be solved by the Numerov method, using the  three-term recurrence
\begin{align}
(1-T_{n+1})\phi_{n+1}-(2+10T_n)\phi_n+(1-T_{n-1})\phi_{n-1}=0, \label{normnum}
\end{align}
where $T_n=-Q_n(\DR)^2/12$, $\DR$ is the integration step length, and the subscript $n$ 
denotes the quantity at the $n$th integration point.
In fact, we employed a slightly modified variant of the method --- the so-called renormalized Numerov \cite{renNum1}.
If we define
\begin{align}
F_n&=(1-T_n)\phi_n,\label{feq}\\
U_n&=\frac{2+10T_n}{1-T_n},
\end{align}
and insert it to Eq. (\ref{normnum}), one multiplication less per step is needed.
The  most  important  point in  the renormalized Numerov method, however, is another substitution
\begin{align}
\mathcal{R}_n&=F_{n+1}/F_n,\label{req}
\end{align}
which leads to a two-term recurrence formula
\begin{align}
\mathcal{R}_n&=U_n-\mathcal{R}_{n-1}^{-1}.\label{renum}
\end{align}
A benefit of Eq. (\ref{renum}) is that $\mathcal{R}_n$ --- in contrast to $F_n$ --- does not grow 
exponentially in the classically forbidden regions \cite{renNum1}.
The initial value equivalent to $\phi_0=0$ and $\phi_1\neq0$ is $\mathcal{R}_0=\infty$ which 
leads to $\mathcal{R}_1=U_1$.
The two-term formula is obtained from a three-term one for the price of forfeiting the information 
about the normalization of the wave function.
However, in this case it is not needed anyway -- only a logarithmic derivative of the function is needed in Eq. (\ref{shifteq3}).
It can be expressed as \cite{renNum1,Blatt}
\begin{align}
&\left[\phi(R)^{-1}d\phi(R)/dR\right]_{R=R_n}\nonumber\\
&=\left(\frac{1/2-T_{n+1}}{1-T_{n+1}}\mathcal{R}_{n}-\frac{1/2-T_{n-1}}{1-T_{n-1}}\mathcal{R}_{n-1}^{-1}\right)\frac{1-T_{n}}{\DR},
\end{align}
and
\begin{align}
&\chi(R_n)^{-1}\left.\frac{d\chi(R)}{dR}\right|_{R=R_n}=\phi(R_n)^{-1}\left.\frac{d\phi(R)}{dR}\right|_{R=R_n}-\frac{1}{2}p(R_n).
\end{align}

As described in Ref. \cite{renNum1}, this method can be also adapted easily to calculate energies 
of the bound states of the system, needed to obtain $B_\mathrm{bound}(T)$ in Eq. 
(\ref{secvirbound}) for helium-4.

In practical application, we chose  $250$ values of the energy $E$, distributed logarithmically in 
the range from $1\times10^{-11}$ to $1$~hartree.
Although the domain of integration in Eq. (\ref{secvirth}) is unbounded, the selected range of 
energies was entirely sufficient, due to the rapidly decaying exponent present in the integrand.
For each value of the energy, we determined $l$ for which the infinite sum in Eq. 
(\ref{ssum}) could be considered converged.
To assess the magnitude of the neglected terms, we used the Born approximation \cite{Joachain}
\begin{align}
\tan   \delta_{l}(E)&\approx -2\mu_\mathrm{a}k \int_0^\infty dR\,j_l^2(kR)V(R)R^2. \label{Bapproxgeneral}
\end{align}
In our case, we assume  the  $-C_6/R^6$ asymptotic behavior of $V(R)$,   which leads to 
\begin{align}
\tan   \delta_{l}(E)&\approx \frac{24\pi\mu_\mathrm{a}^3C_6E^2}{(2l-3)(2l-1)(2l+1)(2l+3)(2l+5)}\label{Bapprox},
\end{align}
where $C_6\approx1.462$ is the total asymptotic coefficient (after summing BO, adiabatic, relativistic, and QED contributions).
The numerical value of $\mu_\mathrm{a}=\mu_\mathrm{n}+1$ was taken from the recent CODATA 18 database 
\cite{codata18} ($3648.149770710(120)$\phantom{.}$m_\mathrm{e}$  for helium-4 and 
$2748.942640035(120)$\phantom{.}$m_\mathrm{e}$ for helium-3).
It must be noted that  our interaction potential includes the retardation  correction 
\cite{Przybytek:12cp,Cencek:12}, which could suggest choosing the  $-C_7/R^7$  long-range form.
However, implementation of the retarding function in Eq. (\ref{Bapproxgeneral}) would be 
cumbersome, so a simultaneously simpler and safer $-C_6/R^6$ assumption was made.

Note that the Born approximation is reliable when $l\gg l_\mathrm{lim}\equiv R_\mathrm{ngl}\sqrt{2\mu_\mathrm{a}E}$, where $R_\mathrm{ngl}$ is the internuclear distance for which the interaction potential can be considered negligible \cite{Joachain}. 
Because of that, the $l$  summation was never stopped below $l_\mathrm{lim}$.
After testing different values, we chose $R_\mathrm{ngl}=150$~bohrs as a safe value for this purpose.

Equation  (\ref{renum}) was propagated separately for each of the $(E,l)$ pairs, with the potential function $V(R)$ described in Sec. \ref{pots} and $\mathcal{W}^\mathrm{int}_{\perp}(R)$, $\mathcal{W}^\mathrm{int}_{\parallel}(R)$, and $V^\mathrm{int}_\mathrm{na}(R)$ taken from Ref. \cite{Przybytek:17}.
%
We followed Ref. \cite{Hurly} in the choice of the integration step $\DR=2\times10^{-5}E^{-1/3}$.
Alternative choices were also tested, but we have observed no significant effect of choosing one 
over the other on the final results.
Equations  (\ref{shifteq1})--(\ref{shifteq3}) were used to calculate the  phase shifts.
As the approximate phase  shifts of   Eqs.  (\ref{shifteq}) and (\ref{shifteq3})   do  not have to be computed at every propagation step, they were tested at every $\lceil 2\pi/(\DR\sqrt{2\mu_\mathrm{a}E})\rceil$th step (i.e., approximately once per wavelength).
The propagation continued until a convergence criterion on the approximate phase  shifts was met.

\begin{figure}[!htb]
\includegraphics[width=1.1\columnwidth]{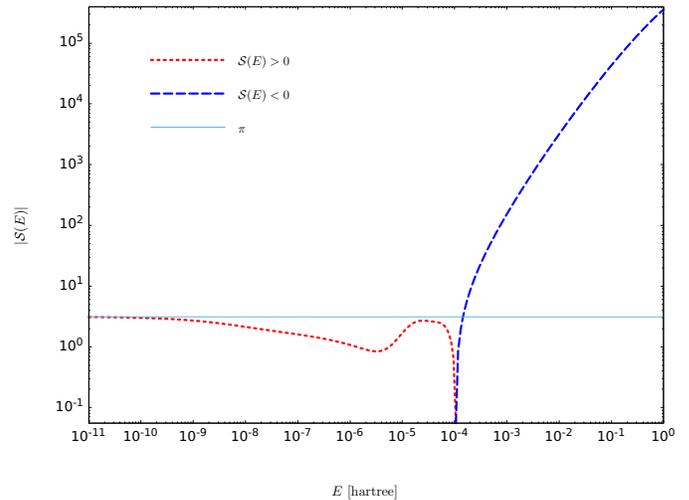}
\caption{Behavior of the function $\mathcal{S}(E)$ for helium-4 in the investigated energy range.}
\label{seplot}
\end{figure}

The shifts were combined with the help of Eq. (\ref{ssum}) to obtain the $\mathcal{S}(E)$ function.
Additionally, the  $\mathcal{S}(0)=\pi$ point was added for helium-4 and $\mathcal{S}(0)=0$ for helium-3, utilizing the Levinson theorem \cite{Joachain}.
Quite interestingly, there are cases in the literature such as Ref. \cite{Kilpatrick}, which use interaction
potentials ``almost'' supporting the bound state of helium-4 and manifest peculiar behavior
of the $\mathcal{S}(E)$ function, which for $E\to 0$ appears to tend to $\pi$ for helium-4, but then rapidly turns
to zero.
Our helium-4 $\mathcal{S}(E)$ curve is presented in Fig. \ref{seplot}.
It is calculated with a potential which undoubtedly supports one bound state of $^4$He$_2$, so it correctly tends to $\pi$.
The calculated $\mathcal{S}(E)$ values for both isotopes can be found in the Supplemental Material  \cite{supp}.



To calculate the second virial coefficient $B(T)$ from Eqs. (\ref{secvir})--(\ref{ssum}), the obtained $\mathcal{S}(E)$ values were interpolated with
 third-order spline functions and numerically integrated, using the \textsc{mathematica} package \cite{Mathematica}.
The calculation was repeated with the uncertainty $\sigma(R)$ of the potential $V(R)$ added or subtracted from it, $V(R)\pm\sigma(R)$, to help estimate the final uncertainty of $B(T)$.
The total $B(T)$ error bar includes three sources of uncertainty, treated as uncorrelated:
\begin{enumerate}[label=(\alph*)]
\item{The first source is the error due to the numerical uncertainty $\sigma(R)$ of the potential $V(R)$, estimated as $(B_+(T)-B_-(T))/2$, where $B_\pm(T)$ denote the $B(T)$ values obtained with $V(R)\pm\sigma(R)$.}
\item{The second source is the error due to interpolation, which was tested in two ways for each $T$:
first, by increasing the order of the interpolating polynomial to $4$, and second, by 
following the method proposed in Ref. \cite{Hurly}, where the authors interpolated $\mathcal{S}(k)$ 
(with $k=\sqrt{2\mu_\mathrm{a}E}$) rather than $\mathcal{S}(E)$, claiming that this method is more 
stable for small $E$. The larger of the two was taken as an estimate of the  interpolation error.
These effects were found to be small, less than $3\%$ of the total uncertainty.}
\item{The third source is the error due to finite accuracy of $\mathcal{S}(E)$  --- less than $1\%$ of the total 
$\sigma$ for $T=1$K, but slowly rising to about $22\%$ of the total error for the highest 
temperature considered. The uncertainty of $\mathcal{S}(E)$ includes both the omitted summation 
terms from Eq. (\ref{ssum}) and the error due to a finite propagation distance during the 
calculation of the phase shifts.}
\end{enumerate}
The potential-related uncertainty (a) dominates in the whole range of temperatures, the other two 
error sources being perceptible only for higher temperatures --- and only because the 
potential-related one decays with $T$ faster.
An analogous error estimation procedure was applied to the acoustic coefficient $\beta_\mathrm{a}(T)$,  Eq.~(\ref{acvir}).

\section{Results and summary}
\begin{table*}[!htb]
\centering
\caption{Second virial coefficient $B(T)$ and second acoustic virial coefficient $\beta_\mathrm{a}(T)$ for $^4$He (in cm$^3$mol$^{-1}$) calculated with our potential ($B_{2020}$, $\beta_{2020}$), compared to the data from Ref. \cite{Cencek:12} ($B_{2012}$, $\beta_{2012}$) for selected temperatures $T$ (in K).}
\begin{adjustbox}{max width=\textwidth}
\begin{tabular}{d{2}d{5}d{3}d{6}d{6}d{2}d{3}d{5}d{3}d{6}d{6}d{2}d{3}}
\hline\hline
$T$   &\centt{$B_{2012}$}& \centt{$\sigma_{2012}$}  &\centt{$B_{2020}$}& \centt{$\sigma_{2020}$}&\centt{$\frac{\sigma_{2012}}{\sigma_{2020}}$}& \centt{$\frac{B_{2020}-B_{2012}}{\sigma_{2012}}$}  &\centt{$\beta_{2012}$}& \centt{$\sigma_{2012}$}  &\centt{$\beta_{2020}$}& \centt{$\sigma_{2020}$}&\centt{$\frac{\sigma_{2012}}{\sigma_{2020}}$}&\centt{$\frac{\beta_{2020}-\beta_{2012}}{\sigma_{2012}}$}\\
\hline
1.00&   -475.74&   0.37&   -475.697&   0.060&   6.2&   12\%&   -536.05&   0.40&   -536.004&   0.067&   6.0&   11\% \\
2.00&   -194.38&   0.13&   -194.369&   0.022&   5.9&   8\%&   -222.35&   0.15&   -222.339&   0.025&   6.0&   8\% \\
5.00&   -64.302&   0.042&   -64.2979&   0.0073&   5.8&   10\%&   -62.979&   0.049&   -62.9744&   0.0085&   5.8&   9\% \\
10.00&   -23.125&   0.020&   -23.1230&   0.0034&   5.9&   10\%&   -13.548&   0.024&   -13.5456&   0.0040&   6.0&   10\% \\
20.00&   -2.7464&   0.0097&   -2.7453&   0.0016&   5.9&   11\%&   10.224&   0.012&   10.2253&   0.0020&   6.1&   11\% \\
30.00&   3.8382&   0.0066&   3.8390&   0.0011&   6.1&   12\%&   17.4638&   0.0083&   17.4649&   0.0013&   6.3&   13\% \\
40.00&   6.9768&   0.0051&   6.97747&   0.00082&   6.3&   13\%&   20.6749&   0.0064&   20.67582&   0.0010&   6.4&   14\% \\
50.00&   8.7506&   0.0041&   8.75111&   0.00066&   6.3&   13\%&   22.3362&   0.0053&   22.33699&   0.00080&   6.6&   15\% \\
100.00&   11.6747&   0.0023&   11.67507&   0.00034&   6.8&   16\%&   24.2708&   0.0030&   24.27138&   0.00042&   7.1&   19\% \\
200.00&   12.1644&   0.0013&   12.16462&   0.00018&   7.3&   17\%&   23.2252&   0.0017&   23.22563&   0.00023&   7.6&   25\% \\
273.15&   11.9279&   0.0010&   11.92814&   0.00013&   7.4&   21\%&   22.2203&   0.0013&   22.22063&   0.00017&   7.6&   25\% \\
300.00&   11.81919&   0.00092&   11.81940&   0.00012&   7.4&   23\%&   21.8763&   0.0012&   21.87663&   0.00016&   7.6&   27\% \\
400.00&   11.40110&   0.00074&   11.401282&   0.000096&   7.7&   25\%&   20.73201&   0.00099&   20.73227&   0.00012&   8.0&   26\% \\
500.00&   11.00715&   0.00062&   11.007306&   0.000079&   7.8&   25\%&   19.77570&   0.00084&   19.77594&   0.00010&   8.1&   28\% \\
1000.00&   9.55038&   0.00037&   9.550487&   0.000045&   8.3&   29\%&   16.62467&   0.00050&   16.624815&   0.000060&   8.4&   29\% \\
\hline\hline                                      
\end{tabular}
\end{adjustbox}
\label{tabres1}
\end{table*}

\begin{table*}[!htb]
\centering
\caption{Second virial coefficient $B(T)$ and second acoustic virial coefficient $\beta_\mathrm{a}(T)$ for $^3$He (in cm$^3$mol$^{-1}$) calculated with our potential ($B_{2020}$, $\beta_{2020}$), compared to the data from Ref. \cite{Cencek:12} ($B_{2012}$, $\beta_{2012}$) for selected temperatures $T$ (in K).}
\begin{adjustbox}{max width=\textwidth}
\begin{tabular}{d{2}d{5}d{3}d{6}d{6}d{2}d{3}d{5}d{3}d{6}d{6}d{2}d{3}}
\hline\hline
$T$   &\centt{$B_{2012}$}& \centt{$\sigma_{2012}$}  &\centt{$B_{2020}$}& \centt{$\sigma_{2020}$}&\centt{$\frac{\sigma_{2012}}{\sigma_{2020}}$}& \centt{$\frac{B_{2020}-B_{2012}}{\sigma_{2012}}$}  &\centt{$\beta_{2012}$}& \centt{$\sigma_{2012}$}  &\centt{$\beta_{2020}$}& \centt{$\sigma_{2020}$}&\centt{$\frac{\sigma_{2012}}{\sigma_{2020}}$}&\centt{$\frac{\beta_{2020}-\beta_{2012}}{\sigma_{2012}}$}\\
\hline
1.00&   -236.39&   0.19&   -236.370&   0.038&   5.0&   10\%&   -299.13&   0.23&   -299.107&   0.044&   5.2&   10\% \\
2.00&   -130.882&   0.094&   -130.871&   0.018&   5.1&   12\%&   -148.87&   0.11&   -148.858&   0.021&   5.2&   11\% \\
5.00&   -47.368&   0.036&   -47.3636&   0.0068&   5.3&   12\%&   -44.550&   0.044&   -44.5453&   0.0080&   5.5&   11\% \\
10.00&   -16.200&   0.018&   -16.1975&   0.0033&   5.5&   14\%&   -6.112&   0.022&   -6.1098&   0.0039&   5.6&   10\% \\
20.00&   0.1061&   0.0093&   0.1071&   0.0016&   5.8&   11\%&   13.281&   0.012&   13.2825&   0.0019&   6.2&   13\% \\
30.00&   5.5362&   0.0064&   5.5370&   0.0011&   6.0&   12\%&   19.2829&   0.0081&   19.2840&   0.0013&   6.2&   13\% \\
40.00&   8.1519&   0.0050&   8.15248&   0.00081&   6.2&   12\%&   21.9338&   0.0063&   21.93471&   0.00099&   6.4&   14\% \\
50.00&   9.6336&   0.0041&   9.63419&   0.00065&   6.3&   14\%&   23.2825&   0.0052&   23.28332&   0.00080&   6.5&   16\% \\
100.00&   12.0385&   0.0022&   12.03883&   0.00034&   6.5&   15\%&   24.6611&   0.0029&   24.66162&   0.00042&   6.9&   18\% \\
200.00&   12.3144&   0.0013&   12.31464&   0.00018&   7.3&   19\%&   23.3863&   0.0017&   23.38665&   0.00023&   7.6&   21\% \\
273.15&   12.02871&   0.00099&   12.02892&   0.00013&   7.4&   21\%&   22.3284&   0.0013&   22.32877&   0.00017&   7.6&   28\% \\
300.00&   11.90860&   0.00092&   11.90880&   0.00012&   7.4&   22\%&   21.9723&   0.0012&   21.97256&   0.00016&   7.6&   22\% \\
400.00&   11.46304&   0.00074&   11.463209&   0.000096&   7.7&   23\%&   20.79843&   0.00099&   20.79870&   0.00012&   8.0&   27\% \\
500.00&   11.05373&   0.00062&   11.053881&   0.000079&   7.8&   24\%&   19.82565&   0.00084&   19.82588&   0.00010&   8.1&   27\% \\
1000.00&   9.56959&   0.00037&   9.569700&   0.000045&   8.3&   30\%&   16.64527&   0.00050&   16.645422&   0.000060&   8.4&   30\% \\
\hline\hline                                      
\end{tabular}
\end{adjustbox}
\label{tabres3}
\end{table*}

In Table \ref{tabres1}, selected  values of the $B(T)$ and $\beta_\mathrm{a}(T)$ coefficients computed by us  for helium-4 are 
 compared to those from Ref. \cite{Cencek:12}.
Table \ref{tabres3} contains analogous data for helium-3. 
Data for more temperature values (including the $0.5$--$1.0$~K range) are presented in the Supplemental Material  \cite{supp}.
The results are in agreement with those of Ref. \cite{Cencek:12}. However, due to the potential of 
a much better quality being used here, the uncertainty of both $B(T)$ and $\beta_\mathrm{a}(T)$ has 
been reduced by a significant factor for the whole investigated temperature range.
The differences between our coefficients and those of Ref. \cite{Cencek:12} do not exceed $31\%$ of the 
estimated $\sigma_{2012}$ uncertainties from Ref. \cite{Cencek:12}, showing that these errors were substantially overestimated.
The possibility of this was pointed out to us by Gao and Pitre, based on their experimental work \cite{Gaoprivate}.
For low temperatures, the changes are below even the more stringent $\sigma_{2020}$.  

As a by-product, the  bound-state energy had to be calculated for helium-4.
It was found to be $-138.88(47)$~neV, confirming the value from the previous calculation \cite{Przybytek:17}.

\begin{figure}[!htb]
\includegraphics[width=1.1\columnwidth]{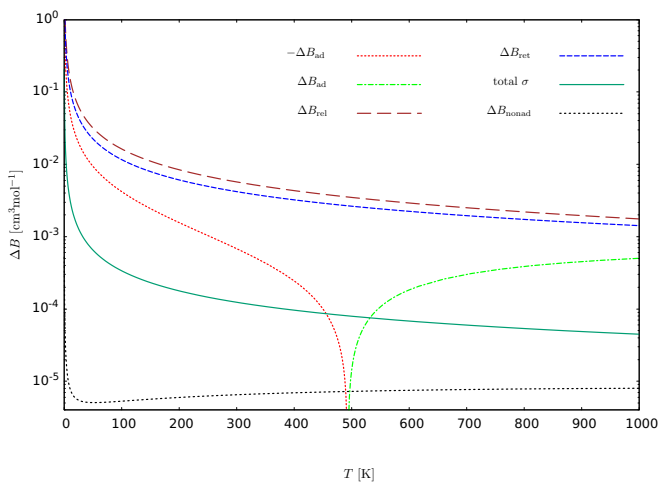}
\caption{Significance of potential contributions to the second virial coefficient $B(T)$ for helium-4 (in cm$^3$mol$^{-1}$)  compared to its uncertainty $\sigma$.}
\label{plotcontr}
\end{figure}


For helium-4, calculations with  particular $V(R)$ contributions turned on or off  were performed to assess the significance of the adiabatic,  retardation  and relativistic potential contributions to $B(T)$.
The  results are presented graphically in Fig. \ref{plotcontr}.
In this context $\Delta B_\mathrm{ad}=B(V_\mathrm{BO}+V_\mathrm{ad})-B(V_\mathrm{BO})$, $\Delta B_\mathrm{rel}=B(V_\mathrm{BO}+V_\mathrm{ad}+V_\mathrm{rel})-B(V_\mathrm{BO}+V_\mathrm{ad})$, $\Delta B_\mathrm{ret}=B(V_\mathrm{ret})-B(V_\mathrm{BO})$ (where $V_\mathrm{ret}$ is $V_\mathrm{BO}$ with the $\sim\!\! 1/R^6$ term retarded \cite{Przybytek:12cp}), and ``total $\sigma$'' is the uncertainty of $B(T)$.
All these values were calculated with Eq. (\ref{BOeq}) with the reduced nuclear mass 
$\mu_\mathrm{n}$ replaced with the reduced mass of two atoms $\mu_\mathrm{a}$.
The plot is consistent with Ref. \cite{Cencek:12}, with the exception that the total uncertainty is now considerably reduced.
Additionally, we tested the validity of using Eq. (\ref{BOeq}) with the atomic reduced mass instead of the nonadiabatic Eq. (\ref{NAPTeq}) in our proper calculations of $B(T)$ with the full potential $V(R)$.
The difference $\Delta B_\mathrm{nonad}=B(V)-B(V)_\mathrm{at}$, where $B(V)_\mathrm{at}$ is the result obtained with Eq. (\ref{BOeq}) with the atomic mass, is also shown in Fig. \ref{plotcontr}.
It is considerably smaller than $\sigma$ in the whole temperature range investigated.
This can be explained by the fact that for helium, when $R$ increases \cite{Przybytek:17},  the functions $\mu_\parallel(R)$ and $\mu_\perp(R)$ quickly reach values close to the reduced mass of two atoms $\mu_\mathrm{a}$.
It can also explain why $\Delta B_\mathrm{nonad}$ rises slightly for higher temperatures: atoms 
with higher kinetic energy are able to penetrate the repulsive part of the interatomic potential 
deeper, where $\mu_\parallel(R)$ and $\mu_\perp(R)$ do have a nontrivial behavior.
However, even for $T=1000$ K this effect is almost six times smaller than the total uncertainty.
Thus, it would be justified to use Eq. (\ref{BOeq}) with the 
atomic masses instead of the more complicated Eq. (\ref{NAPTeq}) not only in this particular case, but probably even more so for heavier atoms, a practice done intuitively before.

\begin{table}[!htb]
\centering
\caption{Second virial coefficient $B(T)$ for $^4$He (in cm$^3$mol$^{-1}$), compared to the experimental data.  Temperature $T$ given in K. The subscript ``$\mathrm{th}$'' denotes our theoretical results,  ``$\mathrm{expt}$''  the  experimental ones, and $\Delta=B^\mathrm{th}-B^\mathrm{expt}$. All the experimental data are calculated by adding the dielectric virial coefficient $b_\varepsilon(T)$ found in Ref. \cite{BoSong} to the measured $B(T)-b_\varepsilon(T)$ from Ref. \cite{Gaiser2020_unpub}, except $T=273.15$K, where $B(T)-b_\varepsilon(T)$ is taken from Ref. \cite{Gaiser2019}.}
\begin{tabular}{d{2}d{7}d{7}d{5}d{5}d{5}}
\hline\hline
$T$   &\centt{$B^\mathrm{th}$}& \centt{$\sigma^\mathrm{th}$}  &\centt{$B^\mathrm{expt}$}& \centt{$\sigma^\mathrm{expt}$}&\centt{$\Delta$}\\
\hline
5.00   & -64.2979&    0.0073&    -64.147 &0.068 & -0.151\\
10.00  & -23.1230&    0.0034&    -23.119 &0.024 & -0.004\\
20.00  &  -2.7453&    0.0016&    -2.734  &0.038 & -0.011\\
30.00  &   3.8390&    0.0011&    3.832   &0.023 &  0.007\\
40.00  &   6.97747&   0.00082&   6.959   &0.019 &  0.018\\
50.00  &   8.75111&   0.00066&   8.732   &0.017 &  0.019\\
100.00 &   11.67507&  0.00034&   11.700  &0.047 & -0.025\\
200.00 &  12.16462&  0.00018&   12.22    &0.21  & -0.06\\
273.15 &  11.92814&  0.00013&   11.9258  &0.0015&  0.0023\\
\hline\hline
\end{tabular}
\label{tabexp}
\end{table}

\begin{table}[!htb]
\centering
\caption{Second acoustic virial coefficient $\beta_\mathrm{a}(T)$ for $^4$He (in cm$^3$mol$^{-1}$), compared to the experimental data from Ref.~\cite{Gavioso2019}.  Temperature $T$ given in K. The subscript ``$\mathrm{th}$'' denotes our theoretical results,  ``$\mathrm{expt}$''  the  experimental ones \cite{Gavioso2019}, and $\Delta=\beta_\mathrm{a}^\mathrm{th}-\beta_\mathrm{a}^\mathrm{expt}$. For   temperatures $273.1600$ and $334.1700$~K, there are several experimental results available.}
\begin{tabular}{d{6}d{7}d{7}d{5}d{5}d{5}}
\hline\hline
$T$   &\centt{$\beta_\mathrm{a}^\mathrm{th}$}& \centt{$\sigma^\mathrm{th}$}  &\centt{$\beta_\mathrm{a}^\mathrm{expt}$}& \centt{$\sigma^\mathrm{expt}$}&\centt{$\Delta$}\\
\hline
235.1400&   22.73307&   0.00020&   22.724&   0.002&   0.009 \\
236.6190&   22.71264&   0.00019&   22.710&   0.003&   0.003 \\
247.0000&   22.57027&   0.00019&   22.566&   0.003&   0.004 \\
260.1200&   22.39314&   0.00018&   22.386&   0.002&   0.007 \\
273.1600&   22.22050&   0.00017&   22.215&   0.005&   0.005 \\
273.1600&   22.22050&   0.00017&   22.216&   0.001&   0.004 \\
273.1600&   22.22050&   0.00017&   22.214&   0.002&   0.006 \\
302.9146&   21.84024&   0.00016&   21.841&   0.004&   -0.001 \\
334.1700&   21.46180&   0.00014&   21.459&   0.002&   0.003 \\
334.1700&   21.46180&   0.00014&   21.460&   0.004&   0.002 \\
362.6000&   21.13590&   0.00013&   21.138&   0.004&   -0.002 \\
395.9000&   20.77519&   0.00013&   20.773&   0.004&   0.002 \\
396.2000&   20.77204&   0.00013&   20.760&   0.005&   0.012 \\
430.2400&   20.42530&   0.00012&   20.416&   0.012&   0.009 \\
\hline\hline
\end{tabular}
\label{tabacexp}
\end{table}

Shortly after submission of this paper, a work \cite{BosongRecent} was published, whose authors also perform calculations of the second virial and the second acoustic virial coefficient for helium-4 and helium-3.
They employ the interatomic interaction potential from Ref. \cite{Przybytek:17} (which our potential expands upon, as described in Sec. III).
Their results for helium-4 agree with ours with respect to the assumed uncertainty, with comparable error bars.
For helium-3 though, there is a substantial disagreement in the whole investigated temperature range.
The authors of Ref. \cite{BosongRecent} themselves note the discrepancy with Ref. \cite{Cencek:12}, which, in turn,  is in perfect agreement with our results, as shown in Table \ref{tabres3}.
It is difficult to pinpoint the cause of this disagreement with certainty.
We note, however, that the potential from Ref. \cite{Przybytek:17} was prepared specifically for helium-4 calculations, and its application to helium-3 requires making certain manual changes in it, namely, rescaling the adiabatic contribution with the correct $^4$He$/^3$He nuclear mass ratio.
There is no information in Ref. \cite{BosongRecent} whether it was corrected in such a way, and if not, it could --- partially at least --- lead to the observed difference in results. This would also explain why the results for helium-4 do agree.
In fact, when neglecting this rescaling, we were able to obtain results for helium-3 consistent with Ref. \cite{BosongRecent} for temperatures greater than $7.5$~K.
The discrepancy for smaller temperatures remains, though, and its source remains unknown.

Although now the second virial coefficient is usually provided by the theory and used to interpret experimental data \cite{Gao1,Gao,Gaiser2020}, not the other way around, there are some recent $B(T)$ measurements available \cite{Gaiser2019,Gaiser2020_unpub}.
Comparison of our $B(T)$ values with those experimental ones is presented in Table \ref{tabexp}.
One should note that the ``experimental'' values in this case are obtained by adding the theoretical second dielectric virial coefficient $b_\varepsilon(T)$ found in Ref. \cite{BoSong} to $B(T)-b_\varepsilon(T)$, which is the actual quantity obtained from the experiment.
Although the dielectric coefficients $b_\varepsilon(T)$ from Ref. \cite{BoSong} have been calculated in a semiclassical approximation only, substituting them with the quantum-statistical results from Refs. \cite{Garberoglio_unpub, Garberoglio_data}, yield no significant change.
They are several orders of magnitude smaller than $B(T)$ and their uncertainty does not contribute 
to the error bar of these values.
Our results agree with the experiment  well, with only two outliers: a 
discrepancy of $2.2\sigma$ for $5$ K and $1.6\sigma$ for $273.15$~K.


Experimental values of the second acoustic virial coefficient $\beta_\mathrm{a}(T)$ for helium-4 can be found in the Supplement  of Ref. \cite{Gavioso2019}.
In Table \ref{tabacexp}, we compare them to our calculations.
The degree of the agreement varies between different temperatures, as well as between different 
measurements for certain $T$.
An explanation of this can be found in Ref. \cite{Gavioso2019} itself: these values of $\beta_\mathrm{a}(T)$ were obtained via a fit to an acoustic model, using all nine cavity modes the measurement was performed for.
On the other hand, the authors noted that the results for some of these acoustic modes are prone to 
errors --- either due to an interference with the elastic resonances of the cavity shell or due to 
an overlap with neighboring modes --- and discarded them from further analysis.
However, as $\beta_\mathrm{a}(T)$ is only an intermediate result in Ref. \cite{Gavioso2019}, it was not recalculated with such a refined data set.
In their previous works, though, the authors of Ref. \cite{Gavioso2019} used such constrained sets of acoustic modes when providing $\beta_\mathrm{a}(T)$, albeit for one temperature only: $\beta_\mathrm{a}(273.16~\mathrm{K})=22.2201(24)$~cm$^3$mol$^{-1}$ \cite{Gavioso2011} and $22.2195(17)$~cm$^3$mol$^{-1}$ \cite{Gavioso2015}.
These results agree perfectly with each other, as well as with our value, $22.22050(17)$~cm$^3$mol$^{-1}$.

Analysis of the results leads to the conclusion that although our potential does not introduce 
any new physical effects if compared to its predecessor \cite{Przybytek:17}, it represents an improvement in the accuracy
and reliability.
The recalculated relativistic and QED components, as well as the augmented set of BO points used, 
not only ensure better justification of the uncertainty estimation but also give us a chance to 
present a bolder, more stringent one.
Hopefully, it should meet the demands of constantly developing experimental metrology in the 
foreseeable future, as well as constitute the next step forward on the path to a new pressure 
standard.
 
\begin{acknowledgments}
We would like to thank Roberto Gavioso, Christof Gaiser, Giovanni Garberoglio and Krzysztof Szalewicz for useful comments and discussions, 
 Bo Gao and Laurent Pitre for pointing out that
our previous error bars for virial coefficients might have been too conservative, and Krzysztof Pachucki for making his server resources available for some of our computations.
This project (Real-K project 18SIB02) has received funding from the EMPIR programme 
cofinanced by the Participating States and from the European Union’s Horizon 2020 research
and innovation programme. The authors also acknowledge support from the National
Science Center, Poland, within Project No. 2017/27/B/ST4/02739.
\end{acknowledgments}


\bibliography{articles}

\begin{thebibliography}{70}
\expandafter\ifx\csname natexlab\endcsname\relax\def\natexlab#1{#1}\fi
\expandafter\ifx\csname bibnamefont\endcsname\relax
  \def\bibnamefont#1{#1}\fi
\expandafter\ifx\csname bibfnamefont\endcsname\relax
  \def\bibfnamefont#1{#1}\fi
\expandafter\ifx\csname citenamefont\endcsname\relax
  \def\citenamefont#1{#1}\fi
\expandafter\ifx\csname url\endcsname\relax
  \def\url#1{\texttt{#1}}\fi
\expandafter\ifx\csname urlprefix\endcsname\relax\def\urlprefix{URL }\fi
\providecommand{\bibinfo}[2]{#2}
\providecommand{\eprint}[2][]{\url{#2}}

\bibitem[{\citenamefont{Gaiser et~al.}(2015)\citenamefont{Gaiser, Zandt, and
  Fellmuth}}]{Gaiser2015}
\bibinfo{author}{\bibfnamefont{C.}~\bibnamefont{Gaiser}},
  \bibinfo{author}{\bibfnamefont{T.}~\bibnamefont{Zandt}}, \bibnamefont{and}
  \bibinfo{author}{\bibfnamefont{B.}~\bibnamefont{Fellmuth}},
  \bibinfo{journal}{Metrologia} \textbf{\bibinfo{volume}{52}},
  \bibinfo{pages}{S217} (\bibinfo{year}{2015}).

\bibitem[{\citenamefont{Gaiser and Fellmuth}(2019)}]{Gaiser2019}
\bibinfo{author}{\bibfnamefont{C.}~\bibnamefont{Gaiser}} \bibnamefont{and}
  \bibinfo{author}{\bibfnamefont{B.}~\bibnamefont{Fellmuth}},
  \bibinfo{journal}{J. Chem. Phys.} \textbf{\bibinfo{volume}{150}},
  \bibinfo{pages}{134303} (\bibinfo{year}{2019}).

\bibitem[{\citenamefont{Moldover et~al.}(2014)\citenamefont{Moldover, Gavioso,
  Mehl, Pitre, de~Podesta, and Zhang}}]{Moldover2014}
\bibinfo{author}{\bibfnamefont{M.~R.} \bibnamefont{Moldover}},
  \bibinfo{author}{\bibfnamefont{R.~M.} \bibnamefont{Gavioso}},
  \bibinfo{author}{\bibfnamefont{J.~B.} \bibnamefont{Mehl}},
  \bibinfo{author}{\bibfnamefont{L.}~\bibnamefont{Pitre}},
  \bibinfo{author}{\bibfnamefont{M.}~\bibnamefont{de~Podesta}},
  \bibnamefont{and} \bibinfo{author}{\bibfnamefont{J.}~\bibnamefont{Zhang}},
  \bibinfo{journal}{Metrologia} \textbf{\bibinfo{volume}{51}},
  \bibinfo{pages}{R1} (\bibinfo{year}{2014}).

\bibitem[{\citenamefont{Gavioso et~al.}(2019)\citenamefont{Gavioso, Ripa,
  Steur, Dematteis, and Imbraguglio}}]{Gavioso2019}
\bibinfo{author}{\bibfnamefont{R.~M.} \bibnamefont{Gavioso}},
  \bibinfo{author}{\bibfnamefont{D.~M.} \bibnamefont{Ripa}},
  \bibinfo{author}{\bibfnamefont{P.~P.~M.} \bibnamefont{Steur}},
  \bibinfo{author}{\bibfnamefont{R.}~\bibnamefont{Dematteis}},
  \bibnamefont{and}
  \bibinfo{author}{\bibfnamefont{D.}~\bibnamefont{Imbraguglio}},
  \bibinfo{journal}{Metrologia} \textbf{\bibinfo{volume}{56}},
  \bibinfo{pages}{045006} (\bibinfo{year}{2019}).

\bibitem[{\citenamefont{Gao et~al.}(2017)\citenamefont{Gao, Pitre, Luo,
  Plimmer, Lin, Zhang, Feng, Chen, and Sparasci}}]{Gao1}
\bibinfo{author}{\bibfnamefont{B.}~\bibnamefont{Gao}},
  \bibinfo{author}{\bibfnamefont{L.}~\bibnamefont{Pitre}},
  \bibinfo{author}{\bibfnamefont{E.~C.} \bibnamefont{Luo}},
  \bibinfo{author}{\bibfnamefont{M.~D.} \bibnamefont{Plimmer}},
  \bibinfo{author}{\bibfnamefont{P.}~\bibnamefont{Lin}},
  \bibinfo{author}{\bibfnamefont{J.~T.} \bibnamefont{Zhang}},
  \bibinfo{author}{\bibfnamefont{X.~J.} \bibnamefont{Feng}},
  \bibinfo{author}{\bibfnamefont{Y.~Y.} \bibnamefont{Chen}}, \bibnamefont{and}
  \bibinfo{author}{\bibfnamefont{F.}~\bibnamefont{Sparasci}},
  \bibinfo{journal}{Measurement} \textbf{\bibinfo{volume}{103}},
  \bibinfo{pages}{258} (\bibinfo{year}{2017}).

\bibitem[{\citenamefont{Gao et~al.}(2020)\citenamefont{Gao, Zhang, Han, Pan,
  Chen, Song, Liu, Hu, Kong, Sparasci et~al.}}]{Gao}
\bibinfo{author}{\bibfnamefont{B.}~\bibnamefont{Gao}},
  \bibinfo{author}{\bibfnamefont{H.~Y.} \bibnamefont{Zhang}},
  \bibinfo{author}{\bibfnamefont{D.~X.} \bibnamefont{Han}},
  \bibinfo{author}{\bibfnamefont{C.~Z.} \bibnamefont{Pan}},
  \bibinfo{author}{\bibfnamefont{H.}~\bibnamefont{Chen}},
  \bibinfo{author}{\bibfnamefont{Y.~N.} \bibnamefont{Song}},
  \bibinfo{author}{\bibfnamefont{W.~J.} \bibnamefont{Liu}},
  \bibinfo{author}{\bibfnamefont{J.~F.} \bibnamefont{Hu}},
  \bibinfo{author}{\bibfnamefont{X.~J.} \bibnamefont{Kong}},
  \bibinfo{author}{\bibfnamefont{F.}~\bibnamefont{Sparasci}},
  \bibnamefont{et~al.}, \bibinfo{journal}{Metrologia, doi:
  10.1088/1681-7575/ab84ca}  (\bibinfo{year}{2020}).

\bibitem[{\citenamefont{Rourke et~al.}(2019)\citenamefont{Rourke, Gaiser, Gao,
  Ripa, and Moldover}}]{Rourke2019}
\bibinfo{author}{\bibfnamefont{P.~M.~C.} \bibnamefont{Rourke}},
  \bibinfo{author}{\bibfnamefont{C.}~\bibnamefont{Gaiser}},
  \bibinfo{author}{\bibfnamefont{B.}~\bibnamefont{Gao}},
  \bibinfo{author}{\bibfnamefont{D.~M.} \bibnamefont{Ripa}}, \bibnamefont{and}
  \bibinfo{author}{\bibfnamefont{M.~R.} \bibnamefont{Moldover}},
  \bibinfo{journal}{Metrologia} \textbf{\bibinfo{volume}{56}},
  \bibinfo{pages}{032001} (\bibinfo{year}{2019}).

\bibitem[{\citenamefont{Schmidt et~al.}(2007)\citenamefont{Schmidt, Gavioso,
  May, and Moldover}}]{Schmidt07}
\bibinfo{author}{\bibfnamefont{J.~W.} \bibnamefont{Schmidt}},
  \bibinfo{author}{\bibfnamefont{R.~M.} \bibnamefont{Gavioso}},
  \bibinfo{author}{\bibfnamefont{E.~F.} \bibnamefont{May}}, \bibnamefont{and}
  \bibinfo{author}{\bibfnamefont{M.~R.} \bibnamefont{Moldover}},
  \bibinfo{journal}{Phys. Rev. Lett.} \textbf{\bibinfo{volume}{98}},
  \bibinfo{pages}{254504} (\bibinfo{year}{2007}).

\bibitem[{\citenamefont{Hendricks}(2018)}]{Hendricks2018}
\bibinfo{author}{\bibfnamefont{J.}~\bibnamefont{Hendricks}},
  \bibinfo{journal}{Nat. Phys.} \textbf{\bibinfo{volume}{14}},
  \bibinfo{pages}{100} (\bibinfo{year}{2018}).

\bibitem[{\citenamefont{Gaiser et~al.}(2020)\citenamefont{Gaiser, Fellmuth, and
  Sabuga}}]{Gaiser2020}
\bibinfo{author}{\bibfnamefont{C.}~\bibnamefont{Gaiser}},
  \bibinfo{author}{\bibfnamefont{B.}~\bibnamefont{Fellmuth}}, \bibnamefont{and}
  \bibinfo{author}{\bibfnamefont{W.}~\bibnamefont{Sabuga}},
  \bibinfo{journal}{Nat. Phys.} \textbf{\bibinfo{volume}{16}},
  \bibinfo{pages}{177} (\bibinfo{year}{2020}).

\bibitem[{\citenamefont{Tiesinga et~al.}()\citenamefont{Tiesinga, Mohr, Newell,
  and Taylor}}]{codata18}
\bibinfo{author}{\bibfnamefont{E.}~\bibnamefont{Tiesinga}},
  \bibinfo{author}{\bibfnamefont{P.~J.} \bibnamefont{Mohr}},
  \bibinfo{author}{\bibfnamefont{D.~B.} \bibnamefont{Newell}},
  \bibnamefont{and} \bibinfo{author}{\bibfnamefont{B.~N.}
  \bibnamefont{Taylor}}, \emph{\bibinfo{title}{{The 2018 CODATA Recommended
  Values of the Fundamental Physical Constants}}}, \bibinfo{howpublished}{(Web
  Version 8.1), http://physics.nist.gov/constants}.

\bibitem[{\citenamefont{Buckingham and Pople}(1955)}]{ClausiusMossotti}
\bibinfo{author}{\bibfnamefont{A.~D.} \bibnamefont{Buckingham}}
  \bibnamefont{and} \bibinfo{author}{\bibfnamefont{J.~A.} \bibnamefont{Pople}},
  \bibinfo{journal}{Trans. Faraday Soc.} \textbf{\bibinfo{volume}{51}},
  \bibinfo{pages}{1029} (\bibinfo{year}{1955}).

\bibitem[{\citenamefont{Kilpatrick}(1953)}]{Kilpaorig}
\bibinfo{author}{\bibfnamefont{J.~E.} \bibnamefont{Kilpatrick}},
  \bibinfo{journal}{J. Chem. Phys.} \textbf{\bibinfo{volume}{21}},
  \bibinfo{pages}{274} (\bibinfo{year}{1953}).

\bibitem[{\citenamefont{Kilpatrick et~al.}(1954)\citenamefont{Kilpatrick,
  Keller, Hammel, and Metropolis}}]{Kilpatrick}
\bibinfo{author}{\bibfnamefont{J.~E.} \bibnamefont{Kilpatrick}},
  \bibinfo{author}{\bibfnamefont{W.~E.} \bibnamefont{Keller}},
  \bibinfo{author}{\bibfnamefont{E.~F.} \bibnamefont{Hammel}},
  \bibnamefont{and}
  \bibinfo{author}{\bibfnamefont{N.}~\bibnamefont{Metropolis}},
  \bibinfo{journal}{Phys. Rev.} \textbf{\bibinfo{volume}{94}},
  \bibinfo{pages}{1103} (\bibinfo{year}{1954}).

\bibitem[{\citenamefont{Hurly and Mehl}(2007)}]{Hurly}
\bibinfo{author}{\bibfnamefont{J.~J.} \bibnamefont{Hurly}} \bibnamefont{and}
  \bibinfo{author}{\bibfnamefont{J.~B.} \bibnamefont{Mehl}},
  \bibinfo{journal}{J. Res. Natl. Inst. Stand. Technol.}
  \textbf{\bibinfo{volume}{112}}, \bibinfo{pages}{75} (\bibinfo{year}{2007}).

\bibitem[{\citenamefont{Pachucki and Komasa}(2008)}]{Pachucki:08}
\bibinfo{author}{\bibfnamefont{K.}~\bibnamefont{Pachucki}} \bibnamefont{and}
  \bibinfo{author}{\bibfnamefont{J.}~\bibnamefont{Komasa}},
  \bibinfo{journal}{J. Chem. Phys.} \textbf{\bibinfo{volume}{129}},
  \bibinfo{pages}{034102} (\bibinfo{year}{2008}).

\bibitem[{\citenamefont{Pachucki and Komasa}(2015)}]{PK2015}
\bibinfo{author}{\bibfnamefont{K.}~\bibnamefont{Pachucki}} \bibnamefont{and}
  \bibinfo{author}{\bibfnamefont{J.}~\bibnamefont{Komasa}},
  \bibinfo{journal}{J. Chem. Phys.} \textbf{\bibinfo{volume}{143}},
  \bibinfo{pages}{034111} (\bibinfo{year}{2015}).

\bibitem[{\citenamefont{Przybytek et~al.}(2017)\citenamefont{Przybytek, Cencek,
  Jeziorski, and Szalewicz}}]{Przybytek:17}
\bibinfo{author}{\bibfnamefont{M.}~\bibnamefont{Przybytek}},
  \bibinfo{author}{\bibfnamefont{W.}~\bibnamefont{Cencek}},
  \bibinfo{author}{\bibfnamefont{B.}~\bibnamefont{Jeziorski}},
  \bibnamefont{and}
  \bibinfo{author}{\bibfnamefont{K.}~\bibnamefont{Szalewicz}},
  \bibinfo{journal}{Phys. Rev. Lett.} \textbf{\bibinfo{volume}{119}},
  \bibinfo{pages}{123401} (\bibinfo{year}{2017}).

\bibitem[{\citenamefont{Joachain}(1975)}]{Joachain}
\bibinfo{author}{\bibfnamefont{C.~J.} \bibnamefont{Joachain}},
  \emph{\bibinfo{title}{Quantum Collision Theory}}
  (\bibinfo{publisher}{North-Holland Publishing Company},
  \bibinfo{year}{1975}).

\bibitem[{\citenamefont{Piel and Chrysos}(2018)}]{Chrysos}
\bibinfo{author}{\bibfnamefont{H.}~\bibnamefont{Piel}} \bibnamefont{and}
  \bibinfo{author}{\bibfnamefont{M.}~\bibnamefont{Chrysos}},
  \bibinfo{journal}{Mol. Phys.} \textbf{\bibinfo{volume}{116}},
  \bibinfo{pages}{2364} (\bibinfo{year}{2018}).

\bibitem[{\citenamefont{Levinson}(1949)}]{Levinson49}
\bibinfo{author}{\bibfnamefont{N.}~\bibnamefont{Levinson}},
  \bibinfo{journal}{Kgl. Danske Videnskab. Selskab. Mat.-fys. Medd.}
  \textbf{\bibinfo{volume}{25}} (\bibinfo{year}{1949}).

\bibitem[{\citenamefont{Cencek et~al.}(2012)\citenamefont{Cencek, Przybytek,
  Komasa, Mehl, Jeziorski, and Szalewicz}}]{Cencek:12}
\bibinfo{author}{\bibfnamefont{W.}~\bibnamefont{Cencek}},
  \bibinfo{author}{\bibfnamefont{M.}~\bibnamefont{Przybytek}},
  \bibinfo{author}{\bibfnamefont{J.}~\bibnamefont{Komasa}},
  \bibinfo{author}{\bibfnamefont{J.~B.} \bibnamefont{Mehl}},
  \bibinfo{author}{\bibfnamefont{B.}~\bibnamefont{Jeziorski}},
  \bibnamefont{and}
  \bibinfo{author}{\bibfnamefont{K.}~\bibnamefont{Szalewicz}},
  \bibinfo{journal}{J. Chem. Phys.} \textbf{\bibinfo{volume}{136}},
  \bibinfo{pages}{224303} (\bibinfo{year}{2012}).

\bibitem[{\citenamefont{Kutzelnigg}(1997)}]{Kutzelnigg:97}
\bibinfo{author}{\bibfnamefont{W.}~\bibnamefont{Kutzelnigg}},
  \bibinfo{journal}{Mol. Phys.} \textbf{\bibinfo{volume}{90}},
  \bibinfo{pages}{909} (\bibinfo{year}{1997}).

\bibitem[{\citenamefont{Handy et~al.}(1986)\citenamefont{Handy, Yamaguchi, and
  {Schaefer III}}}]{Handy:86}
\bibinfo{author}{\bibfnamefont{N.~C.} \bibnamefont{Handy}},
  \bibinfo{author}{\bibfnamefont{Y.}~\bibnamefont{Yamaguchi}},
  \bibnamefont{and} \bibinfo{author}{\bibfnamefont{H.~F.}
  \bibnamefont{{Schaefer III}}}, \bibinfo{journal}{J. Chem. Phys.}
  \textbf{\bibinfo{volume}{84}}, \bibinfo{pages}{4481} (\bibinfo{year}{1986}).

\bibitem[{\citenamefont{Ioannou et~al.}(1996)\citenamefont{Ioannou, Amos, and
  Handy}}]{Ioannou:96}
\bibinfo{author}{\bibfnamefont{A.~G.} \bibnamefont{Ioannou}},
  \bibinfo{author}{\bibfnamefont{R.~D.} \bibnamefont{Amos}}, \bibnamefont{and}
  \bibinfo{author}{\bibfnamefont{N.~C.} \bibnamefont{Handy}},
  \bibinfo{journal}{Chem. Phys. Lett.} \textbf{\bibinfo{volume}{251}},
  \bibinfo{pages}{52 } (\bibinfo{year}{1996}).

\bibitem[{\citenamefont{Handy and Lee}(1996)}]{Handy:96}
\bibinfo{author}{\bibfnamefont{N.~C.} \bibnamefont{Handy}} \bibnamefont{and}
  \bibinfo{author}{\bibfnamefont{A.~M.} \bibnamefont{Lee}},
  \bibinfo{journal}{Chem. Phys. Lett.} \textbf{\bibinfo{volume}{252}},
  \bibinfo{pages}{425 } (\bibinfo{year}{1996}).

\bibitem[{\citenamefont{Gauss et~al.}(2006)\citenamefont{Gauss, Tajti, Kállay,
  Stanton, and Szalay}}]{gauss06}
\bibinfo{author}{\bibfnamefont{J.}~\bibnamefont{Gauss}},
  \bibinfo{author}{\bibfnamefont{A.}~\bibnamefont{Tajti}},
  \bibinfo{author}{\bibfnamefont{M.}~\bibnamefont{Kállay}},
  \bibinfo{author}{\bibfnamefont{J.~F.} \bibnamefont{Stanton}},
  \bibnamefont{and} \bibinfo{author}{\bibfnamefont{P.~G.}
  \bibnamefont{Szalay}}, \bibinfo{journal}{J. Chem. Phys.}
  \textbf{\bibinfo{volume}{125}}, \bibinfo{pages}{144111}
  (\bibinfo{year}{2006}).

\bibitem[{\citenamefont{Tajti et~al.}(2007)\citenamefont{Tajti, Szalay, and
  Gauss}}]{gauss07}
\bibinfo{author}{\bibfnamefont{A.}~\bibnamefont{Tajti}},
  \bibinfo{author}{\bibfnamefont{P.~G.} \bibnamefont{Szalay}},
  \bibnamefont{and} \bibinfo{author}{\bibfnamefont{J.}~\bibnamefont{Gauss}},
  \bibinfo{journal}{J. Chem. Phys.} \textbf{\bibinfo{volume}{127}},
  \bibinfo{pages}{014102} (\bibinfo{year}{2007}).

\bibitem[{\citenamefont{Przybytek et~al.}(2010)\citenamefont{Przybytek, Cencek,
  Komasa, \L{}ach, Jeziorski, and Szalewicz}}]{Przybytek:10}
\bibinfo{author}{\bibfnamefont{M.}~\bibnamefont{Przybytek}},
  \bibinfo{author}{\bibfnamefont{W.}~\bibnamefont{Cencek}},
  \bibinfo{author}{\bibfnamefont{J.}~\bibnamefont{Komasa}},
  \bibinfo{author}{\bibfnamefont{G.}~\bibnamefont{\L{}ach}},
  \bibinfo{author}{\bibfnamefont{B.}~\bibnamefont{Jeziorski}},
  \bibnamefont{and}
  \bibinfo{author}{\bibfnamefont{K.}~\bibnamefont{Szalewicz}},
  \bibinfo{journal}{Phys. Rev. Lett.} \textbf{\bibinfo{volume}{104}},
  \bibinfo{pages}{183003} (\bibinfo{year}{2010}).

\bibitem[{\citenamefont{Bethe and Salpeter}(1957)}]{BeSal}
\bibinfo{author}{\bibfnamefont{H.~A.} \bibnamefont{Bethe}} \bibnamefont{and}
  \bibinfo{author}{\bibfnamefont{E.~E.} \bibnamefont{Salpeter}},
  \emph{\bibinfo{title}{Quantum Mechanics of One- and Two-Electron Atoms}}
  (\bibinfo{publisher}{Springer}, \bibinfo{address}{Berlin},
  \bibinfo{year}{1957}).

\bibitem[{\citenamefont{Cowan and Griffin}(1976)}]{Cowan:76}
\bibinfo{author}{\bibfnamefont{R.~D.} \bibnamefont{Cowan}} \bibnamefont{and}
  \bibinfo{author}{\bibfnamefont{D.~C.} \bibnamefont{Griffin}},
  \bibinfo{journal}{J. Opt. Soc. Am.} \textbf{\bibinfo{volume}{66}},
  \bibinfo{pages}{1010} (\bibinfo{year}{1976}).

\bibitem[{\citenamefont{Araki}(1957)}]{Araki:57}
\bibinfo{author}{\bibfnamefont{H.}~\bibnamefont{Araki}},
  \bibinfo{journal}{Prog. Theor. Phys.} \textbf{\bibinfo{volume}{17}},
  \bibinfo{pages}{619} (\bibinfo{year}{1957}).

\bibitem[{\citenamefont{Sucher}(1958)}]{Sucher:58}
\bibinfo{author}{\bibfnamefont{J.}~\bibnamefont{Sucher}},
  \bibinfo{journal}{Phys. Rev.} \textbf{\bibinfo{volume}{109}},
  \bibinfo{pages}{1010} (\bibinfo{year}{1958}).

\bibitem[{\citenamefont{Pachucki}(1998)}]{Pachucki:98}
\bibinfo{author}{\bibfnamefont{K.}~\bibnamefont{Pachucki}},
  \bibinfo{journal}{J. Phys. B} \textbf{\bibinfo{volume}{31}},
  \bibinfo{pages}{5123} (\bibinfo{year}{1998}).

\bibitem[{\citenamefont{Piszczatowski et~al.}(2009)\citenamefont{Piszczatowski,
  Lach, Przybytek, Komasa, Pachucki, and Jeziorski}}]{Piszcz:09}
\bibinfo{author}{\bibfnamefont{K.}~\bibnamefont{Piszczatowski}},
  \bibinfo{author}{\bibfnamefont{G.}~\bibnamefont{Lach}},
  \bibinfo{author}{\bibfnamefont{M.}~\bibnamefont{Przybytek}},
  \bibinfo{author}{\bibfnamefont{J.}~\bibnamefont{Komasa}},
  \bibinfo{author}{\bibfnamefont{K.}~\bibnamefont{Pachucki}}, \bibnamefont{and}
  \bibinfo{author}{\bibfnamefont{B.}~\bibnamefont{Jeziorski}},
  \bibinfo{journal}{J. Chem. Theory Comput.} \textbf{\bibinfo{volume}{5}},
  \bibinfo{pages}{3039} (\bibinfo{year}{2009}).

\bibitem[{\citenamefont{Korobov}(2004)}]{Korobov:04}
\bibinfo{author}{\bibfnamefont{V.~I.} \bibnamefont{Korobov}},
  \bibinfo{journal}{Phys. Rev. A} \textbf{\bibinfo{volume}{69}},
  \bibinfo{pages}{054501} (\bibinfo{year}{2004}).

\bibitem[{\citenamefont{Drake}(1988)}]{Drake:88}
\bibinfo{author}{\bibfnamefont{G.~W.~F.} \bibnamefont{Drake}},
  \bibinfo{journal}{Nucl. Instr. Meth. Phys. Res. B}
  \textbf{\bibinfo{volume}{31}}, \bibinfo{pages}{7 } (\bibinfo{year}{1988}).

\bibitem[{\citenamefont{Coriani et~al.}(2004)\citenamefont{Coriani, Helgaker,
  J{\o}rgensen, and Klopper}}]{Coriani:04}
\bibinfo{author}{\bibfnamefont{S.}~\bibnamefont{Coriani}},
  \bibinfo{author}{\bibfnamefont{T.}~\bibnamefont{Helgaker}},
  \bibinfo{author}{\bibfnamefont{P.}~\bibnamefont{J{\o}rgensen}},
  \bibnamefont{and} \bibinfo{author}{\bibfnamefont{W.}~\bibnamefont{Klopper}},
  \bibinfo{journal}{J. Chem. Phys.} \textbf{\bibinfo{volume}{121}},
  \bibinfo{pages}{6591} (\bibinfo{year}{2004}).

\bibitem[{\citenamefont{Boys and Bernardi}(1970)}]{Boys:70}
\bibinfo{author}{\bibfnamefont{S.}~\bibnamefont{Boys}} \bibnamefont{and}
  \bibinfo{author}{\bibfnamefont{F.}~\bibnamefont{Bernardi}},
  \bibinfo{journal}{Mol. Phys.} \textbf{\bibinfo{volume}{19}},
  \bibinfo{pages}{553} (\bibinfo{year}{1970}).

\bibitem[{\citenamefont{Aidas et~al.}(2014)\citenamefont{Aidas, Angeli, Bak,
  Bakken, Bast, Boman, Christiansen, Cimiraglia, Coriani, Dahle
  et~al.}}]{daltonpaper}
\bibinfo{author}{\bibfnamefont{K.}~\bibnamefont{Aidas}},
  \bibinfo{author}{\bibfnamefont{C.}~\bibnamefont{Angeli}},
  \bibinfo{author}{\bibfnamefont{K.~L.} \bibnamefont{Bak}},
  \bibinfo{author}{\bibfnamefont{V.}~\bibnamefont{Bakken}},
  \bibinfo{author}{\bibfnamefont{R.}~\bibnamefont{Bast}},
  \bibinfo{author}{\bibfnamefont{L.}~\bibnamefont{Boman}},
  \bibinfo{author}{\bibfnamefont{O.}~\bibnamefont{Christiansen}},
  \bibinfo{author}{\bibfnamefont{R.}~\bibnamefont{Cimiraglia}},
  \bibinfo{author}{\bibfnamefont{S.}~\bibnamefont{Coriani}},
  \bibinfo{author}{\bibfnamefont{P.}~\bibnamefont{Dahle}},
  \bibnamefont{et~al.}, \bibinfo{journal}{WIREs Comput.~Mol.~Sci.}
  \textbf{\bibinfo{volume}{4}}, \bibinfo{pages}{269} (\bibinfo{year}{2014}).

\bibitem[{dal(2013)}]{dalton13}
\emph{\bibinfo{title}{\textsc{dalton}, a molecular electronic structure
  program, release 2013.2}} (\bibinfo{year}{2013}), \bibinfo{note}{see
  http://daltonprogram.org}.

\bibitem[{\citenamefont{Przybytek}(2014)}]{hector}
\bibinfo{author}{\bibfnamefont{M.}~\bibnamefont{Przybytek}}
  (\bibinfo{year}{2014}), \bibinfo{note}{general FCI program \textsc{hector}}.

\bibitem[{dal(2005)}]{dalton2}
\emph{\bibinfo{title}{\textsc{dalton}, a molecular electronic structure
  program, release 2.0}} (\bibinfo{year}{2005}), \bibinfo{note}{see
  http://daltonprogram.org}.

\bibitem[{\citenamefont{Balcerzak et~al.}(2017)\citenamefont{Balcerzak, Lesiuk,
  and Moszynski}}]{Balcerzak:17}
\bibinfo{author}{\bibfnamefont{J.~G.} \bibnamefont{Balcerzak}},
  \bibinfo{author}{\bibfnamefont{M.}~\bibnamefont{Lesiuk}}, \bibnamefont{and}
  \bibinfo{author}{\bibfnamefont{R.}~\bibnamefont{Moszynski}},
  \bibinfo{journal}{Phys. Rev. A} \textbf{\bibinfo{volume}{96}},
  \bibinfo{pages}{052510} (\bibinfo{year}{2017}).

\bibitem[{\citenamefont{Lesiuk and Jeziorski}(2019)}]{lesiuk19}
\bibinfo{author}{\bibfnamefont{M.}~\bibnamefont{Lesiuk}} \bibnamefont{and}
  \bibinfo{author}{\bibfnamefont{B.}~\bibnamefont{Jeziorski}},
  \bibinfo{journal}{J. Chem. Theory Comput.} \textbf{\bibinfo{volume}{15}},
  \bibinfo{pages}{5398} (\bibinfo{year}{2019}).

\bibitem[{\citenamefont{Kutzelnigg}(2008)}]{kutz08}
\bibinfo{author}{\bibfnamefont{W.}~\bibnamefont{Kutzelnigg}},
  \bibinfo{journal}{Int. J. Quantum Chem.} \textbf{\bibinfo{volume}{108}},
  \bibinfo{pages}{2280} (\bibinfo{year}{2008}).

\bibitem[{\citenamefont{Przybytek and Jeziorski}(2012)}]{Przybytek:12ad}
\bibinfo{author}{\bibfnamefont{M.}~\bibnamefont{Przybytek}} \bibnamefont{and}
  \bibinfo{author}{\bibfnamefont{B.}~\bibnamefont{Jeziorski}},
  \bibinfo{journal}{Chem. Phys.} \textbf{\bibinfo{volume}{401}},
  \bibinfo{pages}{170 } (\bibinfo{year}{2012}).

\bibitem[{\citenamefont{Meath and Hirschfelder}(1966)}]{Meath:66}
\bibinfo{author}{\bibfnamefont{W.~J.} \bibnamefont{Meath}} \bibnamefont{and}
  \bibinfo{author}{\bibfnamefont{J.~O.} \bibnamefont{Hirschfelder}},
  \bibinfo{journal}{J. Chem. Phys.} \textbf{\bibinfo{volume}{44}},
  \bibinfo{pages}{3197} (\bibinfo{year}{1966}).

\bibitem[{\citenamefont{Sack}(1964)}]{Sack:64}
\bibinfo{author}{\bibfnamefont{R.~A.} \bibnamefont{Sack}}, \bibinfo{journal}{J.
  Math. Phys.} \textbf{\bibinfo{volume}{5}}, \bibinfo{pages}{260}
  (\bibinfo{year}{1964}).

\bibitem[{\citenamefont{Zhang et~al.}(2006)\citenamefont{Zhang, Yan, Vrinceanu,
  Babb, and Sadeghpour}}]{Zhang:06}
\bibinfo{author}{\bibfnamefont{J.-Y.} \bibnamefont{Zhang}},
  \bibinfo{author}{\bibfnamefont{Z.-C.} \bibnamefont{Yan}},
  \bibinfo{author}{\bibfnamefont{D.}~\bibnamefont{Vrinceanu}},
  \bibinfo{author}{\bibfnamefont{J.~F.} \bibnamefont{Babb}}, \bibnamefont{and}
  \bibinfo{author}{\bibfnamefont{H.~R.} \bibnamefont{Sadeghpour}},
  \bibinfo{journal}{Phys. Rev. A} \textbf{\bibinfo{volume}{74}},
  \bibinfo{pages}{014704} (\bibinfo{year}{2006}).

\bibitem[{\citenamefont{Tang and Toennies}(1984)}]{Tang:84}
\bibinfo{author}{\bibfnamefont{K.~T.} \bibnamefont{Tang}} \bibnamefont{and}
  \bibinfo{author}{\bibfnamefont{J.~P.} \bibnamefont{Toennies}},
  \bibinfo{journal}{J. Chem. Phys.} \textbf{\bibinfo{volume}{80}},
  \bibinfo{pages}{3726} (\bibinfo{year}{1984}).

\bibitem[{\citenamefont{Tang et~al.}(2011)\citenamefont{Tang, Yan, Shi, and
  Mitroy}}]{Tang:11}
\bibinfo{author}{\bibfnamefont{L.-Y.} \bibnamefont{Tang}},
  \bibinfo{author}{\bibfnamefont{Z.-C.} \bibnamefont{Yan}},
  \bibinfo{author}{\bibfnamefont{T.-Y.} \bibnamefont{Shi}}, \bibnamefont{and}
  \bibinfo{author}{\bibfnamefont{J.}~\bibnamefont{Mitroy}},
  \bibinfo{journal}{Phys. Rev. A} \textbf{\bibinfo{volume}{84}},
  \bibinfo{pages}{052502} (\bibinfo{year}{2011}).

\bibitem[{\citenamefont{Przybytek and Jeziorski}(2008)}]{Przybytek:08}
\bibinfo{author}{\bibfnamefont{M.}~\bibnamefont{Przybytek}} \bibnamefont{and}
  \bibinfo{author}{\bibfnamefont{B.}~\bibnamefont{Jeziorski}},
  \bibinfo{journal}{Chem. Phys. Lett.} \textbf{\bibinfo{volume}{459}},
  \bibinfo{pages}{183 } (\bibinfo{year}{2008}).

\bibitem[{\citenamefont{Puchalski et~al.}(2013)\citenamefont{Puchalski, Komasa,
  and Pachucki}}]{Puchalski:13}
\bibinfo{author}{\bibfnamefont{M.}~\bibnamefont{Puchalski}},
  \bibinfo{author}{\bibfnamefont{J.}~\bibnamefont{Komasa}}, \bibnamefont{and}
  \bibinfo{author}{\bibfnamefont{K.}~\bibnamefont{Pachucki}},
  \bibinfo{journal}{Phys. Rev. A} \textbf{\bibinfo{volume}{87}},
  \bibinfo{pages}{030502(R)} (\bibinfo{year}{2013}).

\bibitem[{\citenamefont{Drake}(2006)}]{Handbook}
\bibinfo{editor}{\bibfnamefont{G.~W.~F.} \bibnamefont{Drake}}, ed.,
  \emph{\bibinfo{title}{Handbook of Atomic, Molecular, and Optical Physics}}
  (\bibinfo{publisher}{Springer}, \bibinfo{year}{2006}).

\bibitem[{sup()}]{supp}
\bibinfo{note}{See Supplemental Material for the calculated values of the
  interaction potential contributions, Fortran implementation of the potential,
  values of the virial and acoustic virial coefficients for a larger set of
  temperatures, and calculated values of the $\mathcal{S}(E)$ function.}

\bibitem[{\citenamefont{Przybytek et~al.}(2012)\citenamefont{Przybytek,
  Jeziorski, Cencek, Komasa, Mehl, and Szalewicz}}]{Przybytek:12cp}
\bibinfo{author}{\bibfnamefont{M.}~\bibnamefont{Przybytek}},
  \bibinfo{author}{\bibfnamefont{B.}~\bibnamefont{Jeziorski}},
  \bibinfo{author}{\bibfnamefont{W.}~\bibnamefont{Cencek}},
  \bibinfo{author}{\bibfnamefont{J.}~\bibnamefont{Komasa}},
  \bibinfo{author}{\bibfnamefont{J.~B.} \bibnamefont{Mehl}}, \bibnamefont{and}
  \bibinfo{author}{\bibfnamefont{K.}~\bibnamefont{Szalewicz}},
  \bibinfo{journal}{Phys. Rev. Lett.} \textbf{\bibinfo{volume}{108}},
  \bibinfo{pages}{183201} (\bibinfo{year}{2012}).

\bibitem[{\citenamefont{Numerov}(1923)}]{Numerovorig}
\bibinfo{author}{\bibfnamefont{B.}~\bibnamefont{Numerov}},
  \bibinfo{journal}{Trudy Glavnoi rossiiskoi astrofizicheskoi observatorii}
  \textbf{\bibinfo{volume}{2}}, \bibinfo{pages}{188} (\bibinfo{year}{1923}).

\bibitem[{\citenamefont{Blatt}(1967)}]{Blatt}
\bibinfo{author}{\bibfnamefont{J.~M.} \bibnamefont{Blatt}},
  \bibinfo{journal}{J. Comp. Phys.} \textbf{\bibinfo{volume}{1}},
  \bibinfo{pages}{382} (\bibinfo{year}{1967}).

\bibitem[{\citenamefont{Leroy and Wallace}(1986)}]{Leroy}
\bibinfo{author}{\bibfnamefont{J.~P.} \bibnamefont{Leroy}} \bibnamefont{and}
  \bibinfo{author}{\bibfnamefont{R.}~\bibnamefont{Wallace}},
  \bibinfo{journal}{J. Comput. Phys.} \textbf{\bibinfo{volume}{67}},
  \bibinfo{pages}{239} (\bibinfo{year}{1986}).

\bibitem[{\citenamefont{Johnson}(1977)}]{renNum1}
\bibinfo{author}{\bibfnamefont{B.~R.} \bibnamefont{Johnson}},
  \bibinfo{journal}{J. Chem. Phys.} \textbf{\bibinfo{volume}{67}},
  \bibinfo{pages}{4086} (\bibinfo{year}{1977}).

\bibitem[{Mat()}]{Mathematica}
\bibinfo{note}{Wolfram Research, Inc. \textsc{Mathematica}, Version
  \textsc{7.0.1.0}, Champaign, IL (2009)}.

\bibitem[{Gao()}]{Gaoprivate}
\bibinfo{note}{B. Gao and L. Pitre, private communications (2019)}.

\bibitem[{\citenamefont{Song and Luo}(2020)}]{BoSong}
\bibinfo{author}{\bibfnamefont{B.}~\bibnamefont{Song}} \bibnamefont{and}
  \bibinfo{author}{\bibfnamefont{Q.-Y.} \bibnamefont{Luo}},
  \bibinfo{journal}{Metrologia} \textbf{\bibinfo{volume}{57}},
  \bibinfo{pages}{025007} (\bibinfo{year}{2020}).

\bibitem[{\citenamefont{Gaiser and Fellmuth}(2020)}]{Gaiser2020_unpub}
\bibinfo{author}{\bibfnamefont{C.}~\bibnamefont{Gaiser}} \bibnamefont{and}
  \bibinfo{author}{\bibfnamefont{B.}~\bibnamefont{Fellmuth}},
  \bibinfo{journal}{Metrologia}  (\bibinfo{year}{2020}), \bibinfo{note}{(to be
  published)}.

\bibitem[{\citenamefont{Song et~al.}(2020)\citenamefont{Song, Xu, and
  He}}]{BosongRecent}
\bibinfo{author}{\bibfnamefont{B.}~\bibnamefont{Song}},
  \bibinfo{author}{\bibfnamefont{P.}~\bibnamefont{Xu}}, \bibnamefont{and}
  \bibinfo{author}{\bibfnamefont{M.}~\bibnamefont{He}}, \bibinfo{journal}{Mol.
  Phys., doi: 10.1080/00268976.2020.1802525}  (\bibinfo{year}{2020}).

\bibitem[{\citenamefont{Garberoglio and Harvey}(2020)}]{Garberoglio_unpub}
\bibinfo{author}{\bibfnamefont{G.}~\bibnamefont{Garberoglio}} \bibnamefont{and}
  \bibinfo{author}{\bibfnamefont{A.~H.} \bibnamefont{Harvey}},
  \bibinfo{journal}{J. Res. Natl. Inst. Stand. Technol.}
  \textbf{\bibinfo{volume}{125}}, \bibinfo{pages}{125022}
  (\bibinfo{year}{2020}).

\bibitem[{\citenamefont{Harvey and Garberoglio}(2020)}]{Garberoglio_data}
\bibinfo{author}{\bibfnamefont{A.~H.} \bibnamefont{Harvey}} \bibnamefont{and}
  \bibinfo{author}{\bibfnamefont{G.}~\bibnamefont{Garberoglio}},
  \emph{\bibinfo{title}{Calculated values of the second dielectric and
  refractivity virial coefficients of helium, neon, and argon}}
  (\bibinfo{publisher}{NIST, Gaithersburg, MD}, \bibinfo{year}{2020}),
  \bibinfo{note}{(accessed 2020-07-08)}.

\bibitem[{\citenamefont{Gavioso et~al.}(2011)\citenamefont{Gavioso, Benedetto,
  Ripa, Albo, Guianvarc’h, Merlone, Pitre, Truong, Moro, and
  Cuccaro}}]{Gavioso2011}
\bibinfo{author}{\bibfnamefont{R.~M.} \bibnamefont{Gavioso}},
  \bibinfo{author}{\bibfnamefont{G.}~\bibnamefont{Benedetto}},
  \bibinfo{author}{\bibfnamefont{D.~M.} \bibnamefont{Ripa}},
  \bibinfo{author}{\bibfnamefont{P.~A.~G.} \bibnamefont{Albo}},
  \bibinfo{author}{\bibfnamefont{C.}~\bibnamefont{Guianvarc’h}},
  \bibinfo{author}{\bibfnamefont{A.}~\bibnamefont{Merlone}},
  \bibinfo{author}{\bibfnamefont{L.}~\bibnamefont{Pitre}},
  \bibinfo{author}{\bibfnamefont{D.}~\bibnamefont{Truong}},
  \bibinfo{author}{\bibfnamefont{F.}~\bibnamefont{Moro}}, \bibnamefont{and}
  \bibinfo{author}{\bibfnamefont{R.}~\bibnamefont{Cuccaro}},
  \bibinfo{journal}{Int. J. Thermophys.} \textbf{\bibinfo{volume}{32}},
  \bibinfo{pages}{1339} (\bibinfo{year}{2011}).

\bibitem[{\citenamefont{Gavioso et~al.}(2015)\citenamefont{Gavioso, Ripa,
  Steur, Gaiser, Truong, Guianvarc’h, Tarizzo, Stuart, and
  Dematteis}}]{Gavioso2015}
\bibinfo{author}{\bibfnamefont{R.~M.} \bibnamefont{Gavioso}},
  \bibinfo{author}{\bibfnamefont{D.~M.} \bibnamefont{Ripa}},
  \bibinfo{author}{\bibfnamefont{P.~P.~M.} \bibnamefont{Steur}},
  \bibinfo{author}{\bibfnamefont{C.}~\bibnamefont{Gaiser}},
  \bibinfo{author}{\bibfnamefont{D.}~\bibnamefont{Truong}},
  \bibinfo{author}{\bibfnamefont{C.}~\bibnamefont{Guianvarc’h}},
  \bibinfo{author}{\bibfnamefont{P.}~\bibnamefont{Tarizzo}},
  \bibinfo{author}{\bibfnamefont{F.~M.} \bibnamefont{Stuart}},
  \bibnamefont{and}
  \bibinfo{author}{\bibfnamefont{R.}~\bibnamefont{Dematteis}},
  \bibinfo{journal}{Metrologia} \textbf{\bibinfo{volume}{52}},
  \bibinfo{pages}{S274} (\bibinfo{year}{2015}).

\end{thebibliography}

\end{document}